\documentclass{acta}
\usepackage{supertabular,lscape,epsfig}
\usepackage{amssymb}
\usepackage{amsmath}
\usepackage{graphicx}
\usepackage{longtable}

\SetPages{0}{0}

\SetVol{64}{2014}

\newcommand{\apjl}{{ApJL}}
\newcommand{\nat}{{Nature}}
\newcommand{\iaucirc}{{IAU Circ.}}

\newcommand{\tabmain}{{1}}
\newcommand{\tabIaparam}{{7}}

\newcommand{\figfields}{{1}}
\newcommand{\figfieldssne}{{2}}
\newcommand{\figcadence}{{3}}
\newcommand{\figobsstats}{{4}}
\newcommand{\figSOM}{{5}}
\newcommand{\figLCSNe}{{6}}
\newcommand{\figLCNDN}{{7}}
\newcommand{\figgalfitEx}{{8}}

\newcommand{\figgalfitNohost}{{10}}
\newcommand{\figgalfitRall}{{11}}
\newcommand{\figgalfitRIaCC}{{12}}
\newcommand{\figgalinclination}{{13}}
\newcommand{\figagnsvst}{{14}}
\newcommand{\figLCnuclear}{{15}}
\newcommand{\figpairs}{{16}}
\newcommand{\figIaPrieto}{{17}}
\newcommand{\figIaHubble}{{18}}
\newcommand{\figIaOffset}{{19}}
\newcommand{\figHubbleUnion}{{20}}

\newcommand{\secphotclass}{{3.2}}
\newcommand{\secspecclass}{{3.1}}
\newcommand{\secgalfit}{{4}}

\begin{document}

\begin{Titlepage}
\Title{OGLE-IV Real-Time Transient Search}

\Author{{\L}.~Wyrzykowski$^{1,2}$, 
Z.~Kostrzewa-Rutkowska$^1$, 
S.~Koz{\l}owski$^1$, 
A.~Udalski$^1$, 
R.~Poleski$^{1,3}$,
J.~Skowron$^1$,
N.~Blagorodnova$^2$,
M.~Kubiak$^1$,
M.~K. Szyma{\'n}ski$^1$,
G.~Pietrzy{\'n}ski$^{1,4}$,
I.~Soszy{\'n}ski$^1$,
K.~Ulaczyk$^1$,
P.~Pietrukowicz$^1$,
P.~Mr{\'o}z$^1$
}
{$^1$Warsaw University Observatory\\Al. Ujazdowskie 4, 00-478 Warszawa, Poland\\
(lw,zkostrzewa,simkoz)@astrouw.edu.pl\\
$^2$Institute of Astronomy, University of Cambridge\\Madingley Road, CB3 0HA Cambridge, UK\\
$^3$Department of Astronomy, Ohio State University, 140 W. 18th Ave., Columbus, OH 43210, USA\\
$^4$Universidad de Concepci{\'o}n, Departamento de Astronomia, Casilla 160-C, Concepci{\'o}n, Chile
}

\Received{Month Day, Year}
\end{Titlepage}

\Abstract{
We present the design and first results of a real-time search for transients within the 650 sq. deg. area around the Magellanic Clouds, conducted as part of the OGLE-IV project and aimed at detecting supernovae, novae and other events.
The average sampling of about 4 days from September to May, yielded a detection of 238 transients in 2012/2013 and 2013/2014 seasons.
The superb photometric and astrometric quality of the OGLE data allows for numerous applications of the discovered transients.

We use this sample to prepare and train a Machine Learning-based automated classifier for early light curves, which distinguishes major classes of transients with more than 80\% of correct answers.
Spectroscopically classified 49 supernovae Type Ia are used to construct a Hubble Diagram with statistical scatter of about 0.3 mag and fill the least populated region of the redshifts range in the Union sample.
We investigate the influence of host galaxy environments on supernovae statistics and find the mean host extinction of $A_I$=0.19$\pm$0.10 mag and $A_V$=0.39$\pm$0.21 mag based on a subsample of supernovae Type Ia.
We show that the positional accuracy of the survey is of the order of 0.5 pixels (0.13 arcsec) and that the OGLE-IV Transient Detection System is capable of detecting transients within the nuclei of galaxies.
We present a few interesting cases of nuclear transients of unknown type.

All data on the OGLE transients are made publicly available to the astronomical community via the OGLE website.
}
{surveys, supernovae, novae, transients}

\section{Introduction}
In the last decade wide-field instruments installed on medium-sized telescopes have opened a new window in the time-domain astronomy.
Hundreds of thousands of new variable stars have not only been found (\eg Bramich \etal 2008, Soszy{\'n}ski \etal 2013, Pietrukowicz \etal 2013 , Drake \etal 2014), but also have been well studied thanks to dense photometric coverage collected over many years. 
For example, recently Soszy{\'n}ski \etal (2014) found an RR Lyrae-type star, which over the course of a few years, changed the mode of pulsation from double to single.
Another remarkable example is a merger of a contact binary which resulted in a spectacular explosion (Tylenda \etal 2013).

Long-term sky monitoring programs are also detecting objects, which appear only for a short period of time, \ie transient events. 
Among the recent projects aiming at unbiased large scale observations of large fractions of the sky are SDSS-Stripe82 (Sako \etal 2011), the Catalina Real-Time Transient Survey (CRTS) (Drake \etal 2009), the Palomar Transient Factory (PTF) (Law \etal 2009) and the La Silla Quest (Hadjiyska \etal 2012), to name a few.
In the very near future the ESA's space mission Gaia will provide all-sky detections and classification of transient objects (\eg Wyrzykowski and Hodgkin 2012, Blagorodnova \etal 2014).

Typically, a key focus of most of the transients surveys is on supernovae (SNe), primarily due to their cosmological applications (\eg Riess \etal 1998, Perlmutter \etal 1999, Sullivan \etal 2011, Campbell \etal 2013).
More than 10,000 supernovae has been found to date, however, most of them were hard to study in detail due to lack of good quality photometry and insufficient amount of available spectroscopic follow-up. Another major difficulty is to detect and announce new discoveries as early as possible to allow prompt spectroscopic observations.

Type Ia supernovae, produced in thermonuclear explosion of a white dwarf exceeding a Chandrasekhar's critical mass, are ``standardizable'' candles (\eg Phillips 1993, Prieto \etal 2006) and are now routinely employed in low- and high-redshift studies of the distance scale of the Universe. 
However, the process of  standardization is purely empirical and it still remains unclear which properties of the supernova, its nearest environment and the host are responsible for differences seen in the light curves and spectra of Type Ia SNe. 
The residuals on the Hubble Diagram are often used to study the influence of \eg mass of the host galaxy or its metallicity on the standardization process of the SNe (\eg Childress \etal 2013, Pan \etal 2014) or distribution of extinction in the hosts (Galbany \etal 2012).

Type II supernovae result from a core-collapse (CC) of a massive star.
The range of possible scenarios in those explosions produces a large variety of different subtypes of CC supernovae.
The most common class, Type IIp, characterized by a couple of months long plateau in the light curve, was shown to also be ``standardizable'' and was used as yet another type of distance indicators (\eg Poznanski 2009).

Wide-field and long-term observations increase the number of supernovae of well known types, but also increase chances for detecting rare and unusual examples of supernovae.
Exotic supernovae are being found in both the cores (\eg Mattila \etal 2012), or at the outskirts of galaxies (\eg Maguire \etal 2011), providing additional data for studying the environments of supernovae and their influence on the explosions.
Other new types of transients are also being discovered, such as super-luminous supernovae (SLSN, \eg Quimby \etal 2011) or tidal disruption events (TDE, \eg Gezari \etal 2012).

Here we present the results of the real-time transient search in the first two years of the monitoring program of about 650 sq. deg. around the Magellanic Clouds conducted within the OGLE-IV survey. 
The paper is organized as follows. 
In Section 2 we describe the OGLE-IV survey and the transient detection pipeline. 
Section 3 describes spectroscopic and photometric classifications of the OGLE-IV detections. 
In the next sections we present the applications of OGLE supernovae, studying the influence of environments of supernovae on their properties and light curves (Section 4) and deriving the Hubble Diagram (Section 5).
We conclude in Section 6.

\section{Observations and detection pipeline}

\subsection{The OGLE-IV project}

The Optical Gravitational Lensing Experiment (OGLE) has started in 1992
as one of the first generation microlensing surveys (OGLE-I: 1992-1995).
Since 1997 the OGLE survey started using a new 1.3~m Warsaw Telescope
with a first generation camera (for technical details see Udalski \etal
1997). The OGLE observing facilities are located at the Las Campanas
Observatory, Chile, operated by the Carnegie Institution for Science. 
In 2001 the first generation camera was replaced by an eight detector
CCD mosaic (OGLE-III: 2001-2009, see Udalski \etal 2008a), and in 2010
another instrumental upgrade occured. The newest generation 32 detector
CCD mosaic replaced OGLE-III instrument, marking the start of the
OGLE-IV phase. The new mosaic -- one of the largest CCD cameras
worldwide -- covers the entire field of view of the Warsaw Telescope
(1.4 square degrees). Each CCD is a 4k$\times$2k pixel E2V detector with
$15 \mu$m pixels, giving  the 0.26 arcsec/pixel scale at the focus of
the Warsaw Telescope. OGLE-IV uses only two filters, Johnson-Cousin $I$-
and $V$- bands, however, vast majority of observations are carried out
in the $I$ filter.

OGLE has always been among the largest variability surveys providing
hundreds thousands variable objects of all types and various transients. 
These were primarily microlensing events, found in thousands every year toward
the Galactic Center, which are also used for finding extrasolar planets
(\eg Udalski \etal 2005, Poleski \etal 2014) or studying the structure
of the Galaxy (\eg Wyrzykowski \etal 2014). OGLE has also been
discovering and gathering long-term photometric data for dwarf and
classical novae (\eg Skowron \etal 2009, Mr{\'o}z \etal 2014) and other
rare transient objects (\eg Tylenda \etal 2013).

Since 1994 microlensing events have been searched for in the OGLE data
stream in real-time {\it via} the Early Warning System (EWS, Udalski \etal
1994, Udalski 2003).  Supernovae were first searched for in the
real-time OGLE data in 2003--2004 (NOOS, Udalski 2003, Udalski \etal
2004). However, this project was abandoned because of relatively
small sky coverage during OGLE-III phase and, thus, relatively low
detection rate.

The situation significantly changed in 2010 with the start of OGLE-IV
when the OGLE survey commenced observations of the large area of the sky
around the Magellanic Clouds.  Those fields are dominated by galaxies,
not stars, and thus are perfect for finding extragalactic transients.
The search for transients in the archival OGLE-IV data from years
2010-2012 yielded 130 supernovae and other transients with detailed
well-sampled light curves (Koz{\l}owski \etal 2013). The region around
the South Ecliptic Pole, used for commissioning of the Gaia satellite
mission in 2014, was also searched for archival transients
(Soszy{\'n}ski \etal 2012).

\begin{figure}
\centering
\includegraphics[width=1\columnwidth]{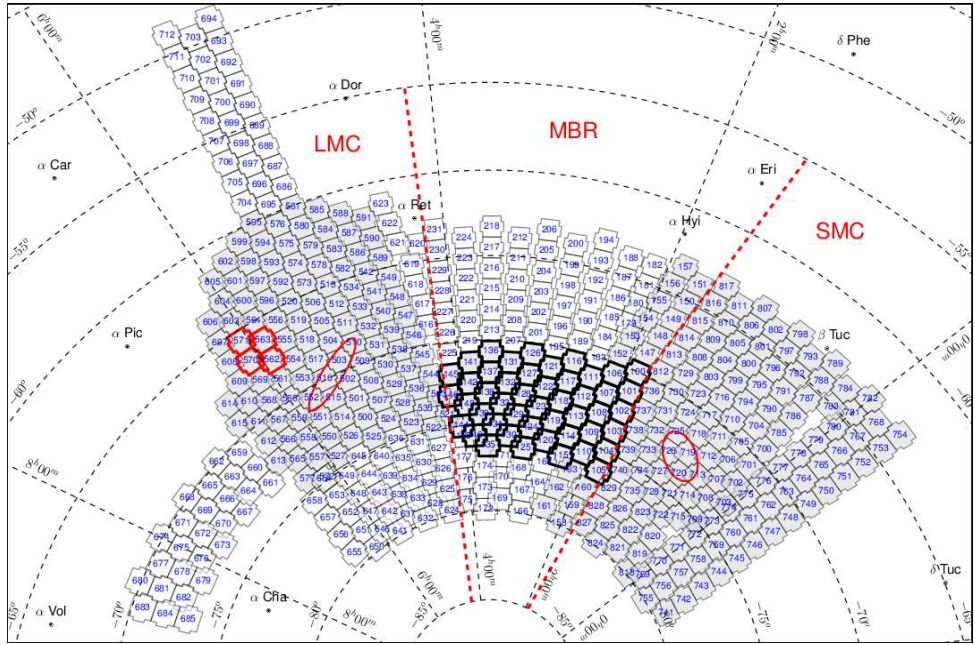}
\caption{Map of the 475 OGLE-IV fields covering about 650 sq.deg., observed since 2010. Grey fields were used for real-time search for supernovae since October 2012 and remaining were added in June 2013. Regions are divided into LMC, MBR and SMC sections. Ellipses mark the positions of the bulk of the stars in the Large and Small Magellanic Cloud. Red-outlined fields are the Gaia SEP fields studied in Soszy{\'n}ski \etal (2012). Black-outlined are the fields searched for transients in Koz{\l}owski \etal (2013).}
\label{fig:fields}
\end{figure}

In October 2012 the real-time processing pipeline was prepared enabling a search for on-going supernovae and other transients in the OGLE-IV data  collected in vicinity of the Magellanic Clouds.
Fig.  \figfields~shows the map of the OGLE-IV fields observed around the Large and Small Magellanic Clouds (LMC and SMC, respectively) and the Magellanic Bridge (MBR) and Fig.  \figfieldssne~displays all detected transients in years 2012--2014. 
Most of the 475 fields have been observed regularly since 2010. 
300 fields were processed in real-time since October 2012 (marked as gray in Fig. \figfields~ and Fig. \figfieldssne) and the remaining 175 were added to the pipeline in June 2013.

\begin{figure}
\centering
\includegraphics[width=1\columnwidth]{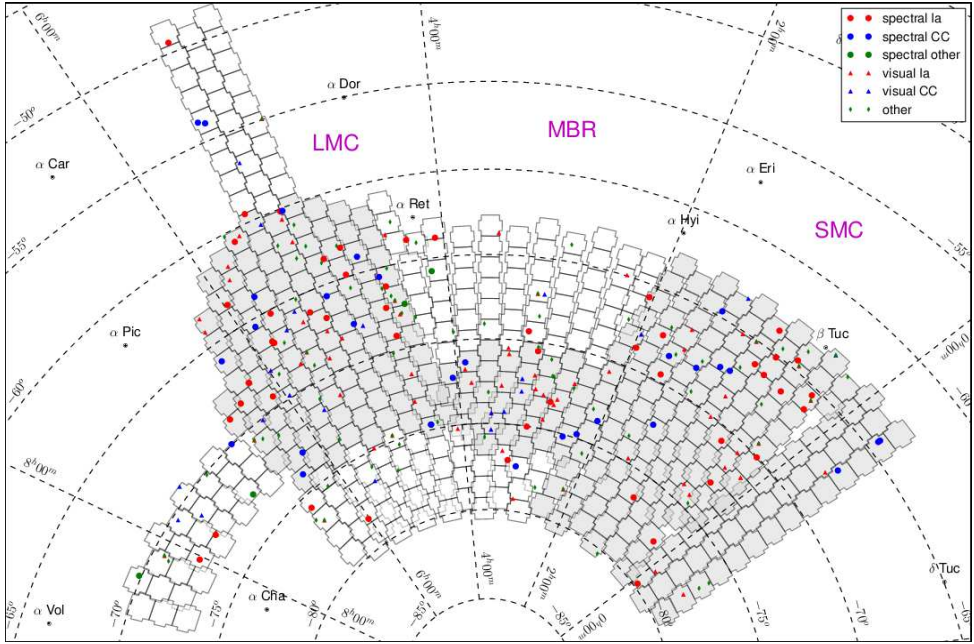}
\caption{OGLE-IV fields as in Fig.\figfields~ displaying positions of all transients discovered in years 2012-2014. Large points show all spectroscopically observed transients, while small dots show the remaining unconfirmed objects, visually classified into classes (See text and Table \tabmain).}
\label{fig:fields-sne}
\end{figure}

\begin{figure}[tb]
\center
\includegraphics[width=13cm]{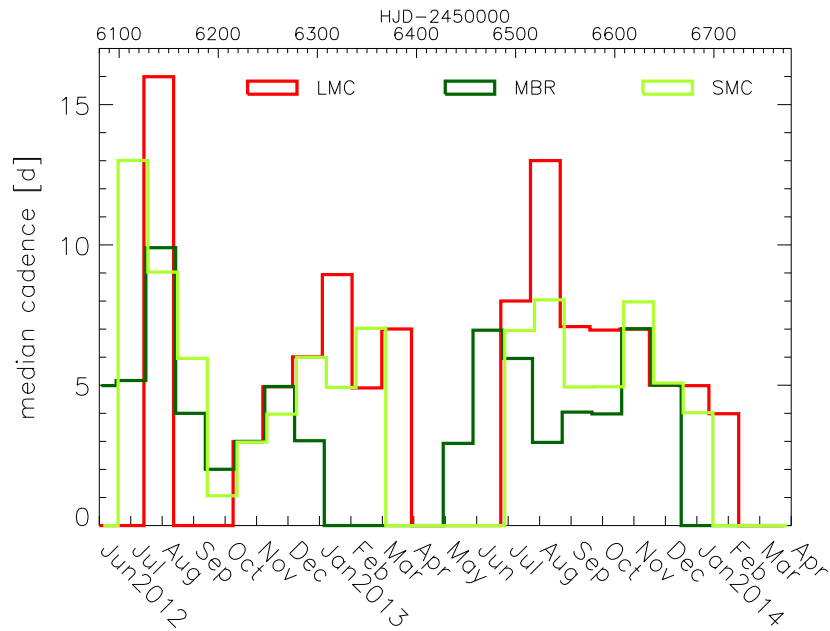}
\caption{Mean monthly cadence between subsequent returning to the same field shown for our three observing regions. Note the cadence increase from 2  to about 4-5 days between two seasons which was caused by the increase in the number of observed fields (See the map in Fig. \figfields).} 
\label{fig:cadence}
\end{figure}

The observing season for the Magellanic Clouds System for the OGLE telescope runs from late July until March, \ie more than eight months, depending on the region of the System.
The region around the SMC is observed for longer period of time that other regions, mostly due to lesser overlap with other OGLE programs in the Bulge and Galactic Disk.
The mean cadence (Fig. \figcadence) also varies through the season, depending of the region. 
The SMC and MBR sections are typically observed with frequency as high as 2 days, whereas the LMC parts can only be observed with 5 days cadence at best.
Note in Fig. \figcadence~that the overall mean cadence has degraded slightly in 2013/2014 seasons, compared to 2012/2013, due to increase in the covered area. 

\begin{figure}[tb]
\center
\includegraphics[width=13cm]{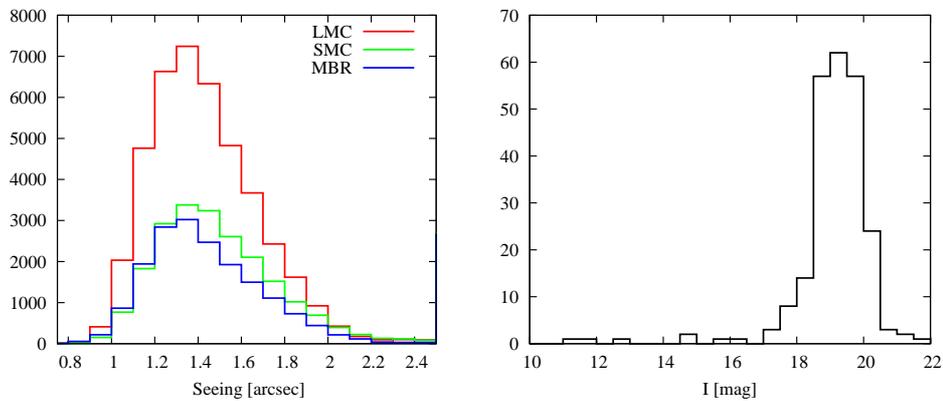}
\caption{
Left panel: distribution of seeing as measured for individual frames in seasons 2012/2013 and 2013/2014 in three different regions (LMC/MBR/SMC). The median seeing was about 1.4 arc sec for all observations. 
Right panel: discovery magnitude of OGLE-IV transients from seasons 2012/2013 and 2013/2014, indicating completeness by $\sim$ 20 mag in $I$-band.
} 
\label{fig:obsstats}
\end{figure}

Fig. \figobsstats~shows in its left panel a distribution of measured
seeing on individual images used in the transient search in seasons
2012/2013 and 2013/2014. The median seeing was about 1.4 arc sec.
There was a significant number of of observations carried under superb seeing condition of about 1 arc second.
The right panel of Fig. \figobsstats~shows a distribution of discovery magnitude of all transients detected in
seasons 2012/2013 and 2013/2014.  
The completeness of detection reaches down to $\sim$ 20 mag, with occasional fainter detections (typically during better seeing conditions).

\subsection{Data reductions}

All the data reduction stages take place at the telescope site in near
real-time. After de-biasing and flat-fielding, the data is processed by
the OGLE real time photometric pipeline (Udalski 2003) which uses the
Difference Imaging Analysis (DIA) technique, fine-tuned to the OGLE data
and using the  Wo{\'z}niak (2000) implementation of Alard and Lupton
(1998) algorithm. The key element in the DIA method is a set of good
quality reference images, which before each subtraction are convolved to
match a given image.

The static database of objects is generated prior to the real-time
processing and uses references images, which are obtained by stacking
several high quality images obtained under excellent seeing conditions
(better than 1 arc second). For the detection of stellar objects and
determination of the reference image fluxes all the reference images
were analyzed with DoPhot (Schechter \etal 1993), which was designed for
PSF photometry of stellar objects. 

In the case of a typical a few arc seconds-wide galaxy, DoPhot
subdivides it into numerous smaller stellar-like objects. However, most
of them are still non-PSF-like, therefore DoPhot flags them as
non-stellar.  For the main database only objects flagged as stars or
likely stars are stored, the remaining are ignored as most of them are
due to various instrumental artifacts or are caused by bright stars. 
Therefore, most galaxies are not in the database and the main search for
transients is performed among objects not matched to template database,
\ie new sources.

The search pipeline is run every day, after the data reduction of the
previous night is finished, typically before Chilean noon.  Then, among
475 fields we select those which were observed last night and in those
data we investigate new objects, returned by the subtraction pipeline.
We select those subtraction residuals which are of positive sign, \ie
are caused by a brightening. The detection threshold for the
brightenings is relatively high to avoid a flood of artifacts. 
The match to Gaussian profile of the PSF profile of brightening must exceed 0.7
(1.0 means perfect match) to classify it as a candidate new object.

In order to assure robust detections and avoid numerous cosmic rays (note, we only take one frame per field during each observing sequence), we also require that the residuals are present on at least two subsequent frames at the same location.
This, therefore, naturally limits possibilities for very early detections of transients while they are still young. 
However, because our sampling is on average 2-5 days (see Fig. \figcadence), our detectability time-frame is still relatively quick.

Another channel for searching for transients is conducted among objects which were there before, but changed their brightness (so called ``old sources'' channel). 
The only difference from the ``new sources'' channel is that the selection is made on all stars from the database which got brighter and remained so over at least two subsequent observations. 
We impose the upper limit on brightness at 18 mag (\ie ignore all brighter objects) in order to avoid numerous variable stars, which are typically brighter in the Magellanic Clouds.

\subsection{Image recognition classifier}

For objects selected from both ``new'' and ``old'' channels, typically about a couple thousand candidates, we generate small cutouts from the subtracted image taken at the brightest epoch. 
Those small imagettes are then fed into an automated image recognition classifier.
The classifier is based on a Self-Organizing Maps (SOM) technique (\eg Wyrzykowski and Belokurov 2008), trained on several thousands of small thumbnail subtraction images. 
The resulting SOM has 5$\times$7 cells and groups in each cell those imagettes which are similar to one another, see Fig. \figSOM.
Note that most of ``yin-yang'' shapes are caused by genuine high proper motion of field stars, see \eg Soszy{\'n}ski \etal (2002), Poleski \etal (2011).
After the training, the cells were visually inspected and labeled according to their thumbnail. 
About 30\% of cells contained images of good subtractions of new stellar-like objects.
The remaining were either bad subtractions, caused by misalignment of images, or effects of high proper motion of stars.
The SOM classifier usually reduced the number of candidates by a factor of two, removing the most obvious and common artifacts.
However, many weirdly shaped artifacts or elongated and twisted cosmic rays still remained as candidates for transients and those were removed at the final stage by visual inspection of both the images and the light curves. 

\begin{figure}
\center
\includegraphics[width=5cm]{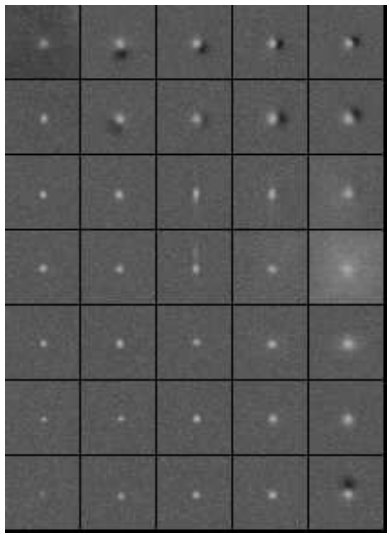}
\caption{Visualization of nodes of the Self-Organizing Map trained on hundreds of cut-outs from the subtracted images. The map had 5$\times$7 nodes/cells. Yin-Yang like images and "shadows" are in most cases visible for stars with high proper motion.}
\label{fig:SOM}
\end{figure}

We have to note here that the OGLE pipeline's new objects selection
criteria are quite stringent, requiring relatively good match to the
Gaussian profile of the PSF. This results in a natural limitation on
detected transients: for an image obtained under typical seeing
condition of Las Campanas Observatory (about 1.2 arc seconds for the
OGLE telescope), the limiting detection magnitude is about 20.0 mag in
the $I$-band. For images taken with exceptionally good seeing (less than
0.9 arc seconds) the detection threshold can go down to 21 mag. However,
the OGLE telescope can reach mag $\sim$22 in a 150~s exposure with
reasonable signal-to-noise. Therefore,  the second stage of the
detection pipeline, which runs on manually selected candidates for
transients (typically a couple per day) we perform a detailed difference
imaging photometry, optimized to the known position of the transient. 
This allows us to obtain better quality photometry, as well as perform
forced photometry of the flux of the transient when it is very faint. 
It also provides the timestamps of non-detection of the transient over
all frames collected for that region, what allows to confirm the
transient nature of the detection.  Because of the data and processing
limitations, we typically run the verification forced photometry of each
transient on frames collected within the same observing season. 

Each candidate's verification photometry is then again inspected
visually and we also check if there is a galaxy visible on the reference
image (\ie more deep than a single frame). Additionally, we perform the
cross-match with the WISE infrared catalog (Wright \etal 2010) in order
to rule out potential transients related to the activity of the galaxy
following the method of Assef \etal (2011) and Koz{\l}owski \etal (2012)
for disentangling between AGNs and galaxies. Given all the above
information we make the final decision about the nature of the
transient.  If it is likely to be a supernova or a classical nova we
give it a name following the pattern: OGLE-year-SN-number, or
OGLE-year-NOVA-number, respectively.  Possible dwarf nova detections or
AGN activity are not reported on discovery, as they are more persistent
than other transients in their nature and, moreover, can be more
effectively searched for in the archival data set (\eg Koz{\l}owski
\etal 2011, Koz{\l}owski \etal 2012, Mowlavi \etal 2014).

All selected on-going transients are made available to the community
{\it via} the web-site updated more or less daily:
\begin{verbatim}
http://ogle.astrouw.edu.pl/ogle4/transients/
\end{verbatim}
For each transient the photometry in the $I$- and $V$-band is provided
as well as the finding chart, subtracted image and the false-color
reference image. The photometry available on the real-time active
web-pages is roughly calibrated to within 0.2 mag, and the calibrated
light curves are obtained after the event is gone and is archived. The
OGLE-IV photometry is tied to precisely calibrated OGLE Photometric Maps
of the Magellanic Clouds (Udalski \etal 2008c) based on observations
collected on hundreds photometric nights. The accuracy of the zero
points of OGLE-IV photometry is better than 0.02~mag.

Transients found by OGLE-IV are also announced via the Astronomers Telegram\footnote{http://www.astronomerstelegram.org}, typically in batches once a week or less often, depending on the number of new detections.

Table \tabmain~provides information on all OGLE-IV transients found by the Transients Detection System in real-time over the period of two seasons 2012/2013 and 2013/2014, with their basic information.
The table presents all transients in order of discovery and contains the following columns:
id of the transient (ID); internal OGLE-IV database id (DBID) in format {\it field.chip.star number}; equatorial coordinates RA$_\mathrm{J2000.0}$ and DEC$_\mathrm{J2000.0}$; 
discovery date as Julian date; 
discovery magnitude in $I$-band; 
Astronomers Telegram number with the discovery (ATEL\#); 
photometric classification  with the highest probability (phot. class., see Section \secphotclass); 
photometric classification probability (prob.); 
spectral type of the transient from the follow-up observations (spec.type, see Section \secspecclass); 
redshift from spectroscopic classification (z) or found in NED\footnote{http://ned.ipac.caltech.edu/} for the nearest galaxy;
ATel number with the spectroscopic classification (ATEL spec.\#) or "NED" if NED redshift available only; 
offset in arc seconds from the nucleus of the host (see Section \secgalfit) or from the nearest galaxy found in NED (in brackets);
offset from the host's center in kpc if redshift available;
offset from the nucleus in units of half-light radius $R_\mathrm{Ser}$ (Offset$^*$, see Section \secgalfit);
half-light radius of the host obtained in {\it galfit} model (see Section \secgalfit);
classification of the nearest galaxy using WISE colors after Assef \etal 2010 (WISE), S=spiral, Ell=elliptical, AGN = active galactic nucleus (see Section \secgalfit);
comment with visual classification type based on the entire light curve.

The data of transients found in years 2012-2014, including fully calibrated photometry and finding charts, are available in the OGLE Archive on the web-site:
\begin{verbatim}
http://ogle.astrouw.edu.pl/ogle4/transients/archive2012-2014
\end{verbatim}

Koz{\l}owski \etal (2013) searched for supernovae and other transients
in the archival OGLE-IV data from years 2010--2012. They also provided
the cumulative number of expected supernovae down to a given peak
$I$-band magnitude using best available rates for most SNe types.
Comparing our real-time detections to those predictions, we estimate
our detection efficiency to be about 50\% down to 19 mag and about
15\% above 20 mag, what is naturally lower than the archival search
(83\% and 38\%, respectively). The lower efficiency is expected, given
the fact that the real-time search relies on much shorter light curves
available at the time of detection, hence only the most robust objects
are typically selected as transients in order to keep the sample as
pure as possible.

\begin{landscape}
\begin{tiny}
\begin{longtable}{l l c c c c c c c c c c c c c c c c}
\hline
ID & DBID & RA$_\mathrm{J2000.0}$ & DEC$_\mathrm{J2000.0}$ & Discovery & Disc. & ATEL & phot. & prob. & spec. & z & ATEL  & Offset & Offset & Offset$^*$ & $R_\mathrm{Ser}$ & WISE & comment \\
OGLE-20.. &  &  &  & HJD-2450000 & $I$ mag & \# & class & & type &  & spec.\#  & [arcsec] & [kpc] & [arcsec] & [arcsec] & & \\
\hline
12-NOVA-02 & SMC714.19.5N & 0:32:55.06 & -74:20:19.7 & 6083.89340 & 11.30 & - & CN & 0.70 & - & - & - & - & - & - & - & - & CN \\
12-NOVA-03 & LMC500.11.11N & 5:20:21.09 & -73:05:43.3 & 6230.75237 & 12.93 & - & CN & 1.00 & - & - & - & - & - & - & - & - & CN \\
12-SN-005 & MBR126.24.184N & 3:00:50.89 & -70:31:11.4 & 6212.79171 & 18.62 & 4481 & Ia & 0.80 & Ia & 0.076 & 4594 & 4.68 & 6.66 & 5.01 & 2.65 & S & Ia \\
12-SN-006 & MBR134.28.15N & 3:33:34.79 & -74:23:40.1 & 6217.74826 & 17.61 & 4495 & Ia & 0.98 & Ibn & 0.06 & 4734 & - & - & - & - & - & Ibn \\
12-SN-007 & MBR124.11.19N & 3:01:13.44 & -75:01:16.1 & 6218.84329 & 18.36 & 4495 & Ia & 1.00 & Ia & 0.059438 & 4602 & - & - & - & - & S & Ia \\
12-SN-008 & LMC537.05.1N & 4:33:37.01 & -71:32:20.4 & 6137.94129 & 18.55 & 4495 & CN & 0.70 & - & - & - & - & - & - & - & Ell & unknown likely Ia \\
12-SN-009 & SMC700.05.8N & 0:13:56.46 & -70:01:41.3 & 6104.93091 & 14.89 & 4495 & Ia & 1.00 & Ia & 0.013966 & CBET3168 & (23.6) & - & - & - & - & Ia \\
12-SN-010 & MBR122.09.9N & 2:58:26.16 & -72:45:12.0 & 6185.78812 & 18.23 & 4495 & IIn & 0.67 & - & - & - & 0.23 & - & 0.25 & 6.15 & S & Ia \\
12-SN-011 & MBR106.20.8N & 2:08:45.14 & -70:39:00.7 & 6141.86720 & 18.52 & 4495 & Ia & 1.00 & - & - & - & 5.11 & - & 5.23 & 2.64 & S & Ia \\
12-SN-012 & LMC535.26.170N & 4:36:17.61 & -73:09:21.9 & 6195.87036 & 18.66 & 4495 & IIn & 0.67 & - & - & - & 4.19 & - & 6.00 & 2.43 & S & Ia \\
12-SN-013 & LMC551.24.8N & 5:42:48.70 & -71:35:52.5 & 6185.88063 & 18.82 & 4495 & IIn & 0.93 & - & - & - & - & - & - & - & S & IIn \\
12-SN-014 & LMC539.06.1N & 4:37:19.18 & -69:08:25.2 & 6150.92144 & 18.20 & 4495 & Ia & 1.00 & Ia & 0.059 & CBET3280 & - & - & - & - & S & Ia \\
12-SN-015 & LMC549.30.12N & 4:33:19.02 & -65:07:56.1 & 6147.94098 & 19.05 & 4495 & CN & 0.70 & - & - & - & - & - & - & - & S & unknown \\
12-SN-016 & LMC562.02.1N & 5:58:56.36 & -68:02:54.9 & 6158.89128 & 19.18 & 4495 & CN & 0.70 & - & - & - & 6.29 & - & 6.59 & 2.55 & S & unknown likely II \\
12-SN-017 & MBR128.21.3N & 3:09:20.61 & -73:00:02.6 & 6111.90423 & 19.13 & 4495 & CN & 0.70 & - & 0.062040 & NED & 10.99 & - & 11.39 & 10.36 & S & IIP or IIl \\
12-SN-018 & SMC717.13.18N & 0:53:37.61 & -70:04:01.0 & 6151.81165 & 19.13 & 4495 & Ia & 0.90 & - & - & - & - & - & - & - & S & Ia \\
12-SN-019 & MBR127.02.6N & 3:09:51.27 & -72:15:24.6 & 6160.90569 & 19.26 & 4495 & Ia & 1.00 & - & - & - & 7.77 & - & 11.56 & 12.36 & S & unknown likely II \\
12-SN-020 & SMC714.22.48N & 0:27:08.75 & -74:14:08.2 & 6140.78721 & 18.50 & 4495 & DN & 1.00 & - & 0.062293 & NED & 6.45 & - & 6.88 & 12.87 & S & Ia \\
12-SN-021 & MBR131.29.12N & 3:21:03.26 & -70:51:13.8 & 6159.90800 & 19.18 & 4495 & Ia & 0.90 & - & - & - & 2.77 & - & 3.59 & 3.88 & S & Ia \\
12-SN-022 & MBR133.05.7N & 3:22:32.00 & -74:16:34.6 & 6159.91212 & 19.76 & 4495 & Ia & 0.90 & - & - & - & - & - & - & - & S & IIP \\
12-SN-023 & LMC554.06.6N & 5:40:58.11 & -68:35:09.5 & 6181.88999 & 19.79 & 4495 & Ia & 1.00 & - & - & - & - & - & - & - & Ell & Ia \\
12-SN-024 & LMC532.20.243N & 4:56:00.97 & -68:00:31.2 & 6230.72111 & 19.49 & 4541 & Ia & 0.32 & - & - & - & - & - & - & - & S & IIP \\
12-SN-025 & MBR127.15.13N & 3:00:35.50 & -71:57:21.3 & 6191.84436 & 19.17 & 4541 & Ia & 1.00 & - & - & - & 0.12 & - & 0.12 & 0.73 & S & Ia \\
12-SN-026 & SMC705.06.76N & 0:24:42.60 & -70:37:07.7 & 6234.71349 & 19.21 & 4578 & IIP & 0.92 & - & - & - & 4.57 & - & 5.71 & 2.60 & S & Ia \\
12-SN-027 & SMC738.03.156N & 1:35:11.24 & -73:35:36.7 & 6242.72138 & 19.30 & 4578 & Ia & 1.00 & I(maybe) & 0.07 & 4558 & - & - & - & - & S & I \\
12-SN-028 & MBR119.09.18N & 2:48:21.42 & -74:26:40.0 & 6237.78721 & 19.75 & 4578 & IIP & 0.92 & - & - & - & - & - & - & - & S & Ia \\
12-SN-029 & MBR132.10.17N & 3:25:33.65 & -72:44:37.8 & 6237.81540 & 19.57 & 4578 & Ia & 0.79 & - & - & - & 0.74 & - & 0.83 & 2.08 & S & Ia \\
12-SN-030 & LMC547.20.71N & 4:31:51.85 & -67:56:47.9 & 6239.84822 & 18.76 & 4578 & Ia & 1.00 & - & 0.066509 & NED & 3.91 & - & 4.18 & 2.08 & S & Ia \\
12-SN-031 & LMC515.15.153N & 5:27:42.45 & -71:15:50.6 & 6243.81787 & 19.87 & 4578 & Ia & 1.00 & - & - & - & - & - & - & - & S & Ia \\
12-SN-032 & LMC506.02.16N & 5:22:08.05 & -66:09:47.2 & 6248.81761 & 18.73 & 4578 & Ia & 1.00 & Ia & 0.063 & 4602 & - & - & - & - & Ell & Ia \\
12-SN-033 & LMC555.12.119N & 5:43:19.40 & -67:04:21.0 & 6245.78759 & 20.09 & 4578 & Ia & 1.00 & - & - & - & - & - & - & - & - & unknown \\
12-SN-034 & LMC522.23.9N & 4:21:17.75 & -74:38:04.7 & 6246.78692 & 19.93 & 4604 & IIP & 0.70 & - & - & - & - & - & - & - & S & unknown \\
12-SN-035 & LMC574.23.5N & 5:16:44.86 & -62:59:03.5 & 6253.81142 & 19.48 & 4604 & IIP & 0.48 & - & - & - & 6.60 & - & 6.67 & 4.19 & S & unknown \\
12-SN-036 & LMC606.01.2N & 6:15:42.01 & -64:08:57.9 & 6247.82342 & 18.75 & 4604 & IIP & 0.48 & - & - & - & 3.77 & - & 7.00 & 4.71 & S & Ia \\
12-SN-037 & LMC611.14.4N & 6:21:27.72 & -70:05:48.5 & 6158.91761 & 19.56 & 4604 & CN & 0.70 & - & - & - & 5.07 & 3.59 & 7.73 & 7.05 & S & unknown likely II \\
12-SN-038 & LMC561.22.79N & 5:55:53.05 & -68:28:43.5 & 6249.76224 & 19.96 & 4604 & IIP & 0.70 & - & - & - & 8.78 & - & 10.35 & 7.21 & S & unknown \\
12-SN-039 & LMC612.20.18N & 6:29:40.49 & -70:55:36.0 & 6252.84831 & 20.02 & 4604 & IIP & 0.70 & - & - & - & 4.06 & - & 5.82 & 2.74 & S & unknown \\
12-SN-040 & LMC568.14.15N & 6:07:01.59 & -69:21:17.1 & 6264.81780 & 15.95 & 4604 & Ia & 1.00 & Ia & 0.014690 & 4618 & 0.76 & 0.22 & 0.76 & 3.93 & Ell & Ia \\
12-SN-041 & LMC538.10.26N & 4:42:23.37 & -70:05:46.3 & 6264.76484 & 19.79 & 4641 & Ia & 0.30 & - & - & - & - & - & - & - & S & unknown \\
12-SN-042 & LMC595.07.4N & 5:25:27.44 & -60:24:35.8 & 6246.76639 & 18.66 & 4641 & IIn & 0.67 & - & - & - & 7.52 & - & 7.52 & 2.71 & S & unknown likely II \\
12-SN-043 & LMC576.24.6N & 5:15:38.11 & -60:36:06.4 & 6266.82530 & 19.07 & 4641 & Ia & 0.99 & - & - & - & - & - & - & - & S & Ia \\
12-SN-044 & LMC576.32.362N & 5:14:53.48 & -60:08:47.7 & 6276.79792 & 18.42 & 4641 & Ia & 1.00 & Ia & 0.06 & 4734 & - & - & - & - & S & Ia \\
12-SN-045 & LMC580.19.9N & 5:11:28.24 & -61:08:29.7 & 6285.75809 & 19.66 & 4657 & Ia & 0.90 & - & - & - & - & - & - & - & - & unknown \\
12-SN-046 & LMC595.28.7N & 5:29:47.99 & -59:29:59.4 & 6277.82714 & 19.27 & 4657 & Ia & 0.90 & Ia & 0.111 & 4734 & 1.69 & 3.37 & 2.53 & 0.91 & S & Ia \\
12-SN-047 & MBR144.11.34N & 4:04:14.25 & -75:12:08.2 & 6263.72945 & 19.67 & 4657 & Ia & 0.36 & - & - & - & 1.16 & - & 1.51 & 0.76 & S & Ia \\
12-SN-048 & SMC732.19.38N & 1:23:14.23 & -72:19:54.7 & 6297.58574 & 18.37 & 4676 & IIn & 0.67 & IIn & 0.067 & 4696 & 3.58 & 4.54 & 3.82 & 3.88 & S & IIn \\
\hline
\caption{Properties of all transients found by the OGLE-IV Transients Detection System in seasons 2012-2014. See main text for details.}
\label{tab:main}
\end{longtable}
\end{tiny}
\end{landscape}
\begin{landscape}
\begin{tiny}
\begin{longtable}{l l c c c c c c c c c c c c c c c c}
\hline
ID & DBID & RA$_\mathrm{J2000.0}$ & DEC$_\mathrm{J2000.0}$ & Discovery & Disc. & ATEL & phot. & prob. & spec. & z & ATEL  & Offset & Offset & Offset$^*$ & $R_\mathrm{Ser}$ & WISE & comment \\
 &  &  &  & HJD-2450000 & $I$ mag & \# & class & & type &  & spec.\#  & [arcsec] & [kpc] & [arcsec] & [arcsec] & & \\
\hline
12-SN-049 & SMC711.14.32N & 0:37:48.09 & -70:44:46.5 & 6285.58220 & 18.65 & 4676 & IIn & 0.67 & Ia & 0.07 & 4696 & - & - & - & - & S & Ia \\
12-SN-050 & LMC564.04.23N & 5:52:34.36 & -65:26:59.7 & 6311.71617 & 18.19 & 4689 & IIn & 0.97 & IIn & 0.082 & 4696 & 3.56 & 5.43 & 3.58 & 2.04 & S & IIn \\
12-SN-051 & LMC578.24.12N & 5:04:06.71 & -63:34:57.0 & 6290.78024 & 19.73 & 4689 & Ia & 0.90 & Ia & 0.11 & 4698 & 6.32 & 12.52 & 6.90 & 2.22 & S & Ia \\
12-SN-052 & LMC599.23.6N & 5:36:29.34 & -60:28:37.5 & 6255.80414 & 18.97 & 4690 & IIP & 0.48 & - & - & - & 0.28 & - & 0.42 & 3.62 & S & Ia \\
13-SN-001 & SMC797.03.426N & 0:46:30.44 & -65:08:16.4 & 6297.55517 & 18.83 & 4718 & IIP & 0.48 & Ia & 0.09 & 4721 & - & - & - & - & S & Ia \\
13-SN-002 & SMC810.26.1800N & 1:23:30.46 & -64:39:42.9 & 6301.57563 & 18.52 & 4718 & IIn & 0.67 & II & 0.0722 & 4721 & 1.10 & 1.49 & 1.17 & 1.72 & S & IIn \\
13-SN-003 & SMC743.13.16N & 23:15:46.39 & -76:58:17.1 & 6263.59832 & 18.43 & 4718 & Ia & 0.46 & - & - & - & 0.09 & - & 0.10 & 3.47 & S & unknown likely nuclear \\
13-SN-004 & SMC769.06.26N & 23:53:52.50 & -79:46:19.0 & 6287.59400 & 18.61 & 4718 & Ia & 1.00 & Ia & 0.06 & 4734 & 13.20 & 15.11 & 13.89 & 4.62 & S & Ia \\
13-SN-005 & SMC799.18.93N & 1:01:37.14 & -67:18:01.7 & 6258.59090 & 19.68 & 4718 & IIP & 0.48 & II & 0.07 & 4734 & - & - & - & - & S & IIP \\
13-SN-006 & SMC800.03.333N & 0:58:08.06 & -66:48:26.7 & 6271.54521 & 19.44 & 4718 & Ia & 0.90 & - & 0.069548 & NED & (24.7) & - & - & - & - & unknown \\
13-SN-007 & SMC805.29.180N & 1:07:56.84 & -65:09:14.4 & 6290.63757 & 19.24 & 4718 & IIP & 0.92 & - & - & - & 16.64 & - & 16.72 & 5.32 & - & unknown likely Ia \\
13-SN-008 & SMC813.32.1389N & 1:30:57.95 & -67:39:12.4 & 6237.66789 & 18.82 & 4718 & Ia & 1.00 & - & - & - & 2.63 & - & 4.25 & 2.85 & S & Ia \\
13-SN-009 & LMC549.03.218N & 4:36:08.21 & -66:09:41.2 & 6296.74994 & 18.67 & 4746 & Ia & 1.00 & Ia & 0.060 & 4734 & 2.20 & 2.52 & 2.20 & 2.79 & S & Ia \\
13-SN-010 & MBR142.13.82N & 3:49:19.65 & -72:33:34.6 & 6309.61450 & 19.26 & 4746 & IIP & 0.48 & - & - & - & - & - & - & - & S & Ia \\
13-SN-011 & LMC586.12.12N & 4:47:10.92 & -64:02:36.5 & 6319.65283 & 18.83 & 4746 & IIP & 0.48 & IIP & 0.05 & 4754 & 5.00 & 4.82 & 5.04 & 3.07 & S & IIP \\
13-SN-012 & LMC512.27.28N & 5:09:52.38 & -65:39:48.6 & 6319.62598 & 19.07 & - & IIn & 0.93 & IIn & 0.08 & 4774 & - & - & - & - & Ell & IIn \\
13-SN-013 & MBR141.02.36N & 3:53:38.01 & -71:45:32.3 & 6309.61246 & 19.15 & 4825 & Ia & 1.00 & - & - & - & - & - & - & - & - & Ia \\
13-SN-014 & MBR141.11.46N & 3:53:07.78 & -71:20:46.6 & 6334.61410 & 17.60 & 4825 & Ia & 0.80 & II & 0.037 & 4829 & 1.03 & 0.75 & 1.11 & 2.35 & S & IIP \\
13-SN-015 & MBR155.18.157N & 2:02:21.55 & -65:44:08.4 & 6338.53517 & 19.38 & 4825 & Ia & 1.00 & Ia & 0.088 & 4829 & - & - & - & - & S & Ia \\
13-SN-016 & LMC581.08.1389 & 5:13:52.76 & -60:09:26.9 & 6379.55018 & 18.68 & 4859 & IIP & 1.00 & IIn & 0.074 & 4860 & - & - & - & - & AGN & IIn \\
13-SN-017 & LMC522.21.1824 & 4:27:22.20 & -74:42:09.4 & 6352.58671 & 18.27 & 4859 & DN & 1.00 & IIn & 0.043 & 4863 & 3.44 & 2.87 & 4.45 & 5.66 & S & IIn \\
13-SN-018 & LMC540.23.45N & 4:39:13.81 & -67:23:36.4 & 6354.57519 & 18.98 & 4875 & Ia & 0.36 & Ia & 0.067 & 4880 & 0.14 & 0.17 & 0.21 & 2.12 & S & Ia \\
13-SN-019 & MBR114.11.509 & 2:27:33.91 & -75:08:16.0 & 6363.50225 & 19.48 & 4875 & Ia & 0.94 & IIn & 0.079 & 4876 & - & - & - & - & S & IIP \\
13-SN-020 & LMC575.15.23N & 5:15:17.87 & -62:00:44.5 & 6354.62311 & 19.51 & 4917 & Ia & 1.00 & - & - & - & 1.94 & - & 3.31 & 3.59 & S & Ia \\
13-SN-021 & LMC519.14.219N & 5:28:01.18 & -66:30:09.0 & 6363.54948 & 18.71 & 4917 & Ia & 0.90 & - & - & - & - & - & - & - & Ell & Ia \\
13-SN-022 & MBR121.29.878 & 2:50:17.36 & -70:52:27.8 & 6369.51506 & 20.01 & 4917 & Ia & 0.98 & - & - & - & 3.30 & - & 4.55 & 2.65 & S & unknown \\
13-SN-023 & LMC552.07.190N & 5:41:43.68 & -70:56:22.7 & 6375.58726 & 19.82 & 4917 & Ia & 0.98 & - & - & - & - & - & - & - & - & unknown \\
13-SN-024 & LMC510.26.4702N & 5:10:48.57 & -68:16:23.8 & 6347.59765 & 19.50 & 4917 & Ia & 1.00 & - & - & - & - & - & - & - & S & Ia \\
13-SN-025 & LMC606.27.276 & 6:13:53.98 & -63:21:25.4 & 6389.55255 & 18.80 & 5000 & Ia & 0.90 & - & - & - & 0.19 & - & 0.19 & 4.66 & S & unknown likely Ia \\
13-SN-026 & SMC723.10.22N & 1:13:38.51 & -69:32:11.4 & 6438.86809 & 19.26 & 5166 & IIP & 0.48 & - & - & - & 1.52 & - & 2.50 & 1.31 & S & unknown \\
13-SN-027 & SMC714.14.62N & 0:22:59.80 & -74:38:32.3 & 6459.94906 & 18.70 & 5166 & IIP & 0.48 & - & - & - & 0.14 & - & 0.16 & 1.69 & S & unknown \\
13-SN-028 & SMC789.27.52N & 0:25:34.22 & -62:54:52.2 & 6438.90375 & 19.73 & 5166 & IIP & 0.48 & - & 0.034300 & NED & 4.10 & - & 4.92 & 7.50 & S & unknown likely II \\
13-SN-029 & SMC714.26.7746 & 0:36:16.30 & -73:59:57.7 & 6467.85673 & 18.77 & 5166 & IIP & 0.92 & - & - & - & 3.20 & - & 3.55 & 3.48 & S & unknown likely Ia \\
13-SN-030 & SMC814.30.44N & 1:32:41.37 & -66:25:43.3 & 6464.93131 & 18.83 & 5204 & IIP & 0.48 & - & - & - & 0.73 & - & 0.93 & 0.72 & - & Ia \\
13-SN-031 & SMC813.22.123 & 1:33:51.35 & -68:07:57.5 & 6464.92926 & 19.68 & 5204 & CN & 0.70 & - & - & - & - & - & - & - & S & unknown \\
13-SN-032 & SMC787.30.67N & 0:19:17.49 & -65:24:42.6 & 6480.93958 & 19.26 & 5204 & Ia & 0.90 & Ia & 0.09 & 5233 & 10.20 & 16.91 & 10.50 & 2.80 & S & Ia \\
13-SN-033 & SMC743.30.2N & 23:16:01.84 & -76:21:09.9 & 6474.90090 & 19.07 & 5204 & IIP & 0.45 & - & - & - & 0.17 & - & 0.18 & 4.14 & S & unknown nuclear \\
13-SN-034 & SMC702.29.613 & 0:10:47.75 & -71:28:28.2 & 6482.92392 & 19.68 & 5204 & Ia & 0.90 & - & - & - & 1.62 & - & 1.78 & 7.90 & S & Ia \\
13-SN-035 & MBR153.24.65N & 1:58:24.30 & -67:55:54.5 & 6479.95207 & 18.91 & 5220 & IIP & 0.48 & - & - & - & 0.91 & - & 1.84 & 2.45 & S & unknown likely II \\
13-SN-036 & SMC716.22.274N & 0:54:27.93 & -68:37:30.7 & 6483.85725 & 19.08 & 5220 & Ia & 0.36 & - & - & - & 1.76 & - & 2.25 & 2.33 & S & Ia \\
13-SN-037 & SMC808.14.42N & 1:18:00.44 & -67:43:16.3 & 6487.90527 & 19.88 & 5220 & IIn & 0.53 & - & - & - & 0.09 & - & 0.12 & 1.84 & S & AGN-flare \\
13-SN-038 & SMC751.09.55N & 23:39:24.77 & -67:02:54.8 & 6494.70677 & 17.97 & 5235 & IIn & 0.67 & II & 0.045 & 5249 & - & - & - & - & S & unknown \\
13-SN-039 & SMC707.04.24N & 0:22:33.02 & -72:53:41.9 & 6494.76434 & 18.94 & 5235 & Ia & 0.90 & Ia & 0.08 & 5233 & 2.22 & 3.31 & 3.10 & 3.47 & S & Ia \\
13-SN-040 & SMC802.23.63N & 0:53:49.35 & -63:49:24.1 & 6508.81760 & 19.25 & 5251 & Ia & 0.80 & Ia & 0.09 & 5250 & 1.22 & 2.03 & 1.49 & 1.41 & S & Ia \\
13-SN-041 & SMC812.13.110N & 1:36:17.96 & -69:34:12.5 & 6508.83856 & 19.44 & 5251 & Ia & 1.00 & Ia & 0.10 & 5250 & 2.39 & 4.36 & 2.42 & 3.35 & S & Ia \\
13-SN-042 & SMC706.06.7602 & 0:23:16.85 & -71:42:49.2 & 6507.85177 & 19.82 & 5251 & Ia & 1.00 & - & - & - & 0.69 & - & 0.69 & 1.37 & S & Ia \\
\hline
\caption{Properties of all transients found by the OGLE-IV Transients Detection System in seasons 2012-2014. See main text for details.}
\end{longtable}
\end{tiny}
\end{landscape}
\begin{landscape}
\begin{tiny}
\begin{longtable}{l l c c c c c c c c c c c c c c c c}
\hline
ID & DBID & RA$_\mathrm{J2000.0}$ & DEC$_\mathrm{J2000.0}$ & Discovery & Disc. & ATEL & phot. & prob. & spec. & z & ATEL  & Offset & Offset & Offset$^*$ & $R_\mathrm{Ser}$ & WISE & comment \\
 &  &  &  & HJD-2450000 & $I$ mag & \# & class & & type &  & spec.\#  & [arcsec] & [kpc] & [arcsec] & [arcsec] & & \\
\hline
13-SN-043 & SMC800.23.1047 & 0:52:48.79 & -66:09:48.6 & 6508.81269 & 20.27 & 5251 & IIP & 0.70 & Ia & 0.14 & 5290 & 2.29 & 5.60 & 2.49 & 2.02 & S & Ia \\
13-SN-044 & SMC820.14.61N & 0:18:50.31 & -77:33:32.6 & 6511.78989 & 18.75 & 5277 & Ia & 0.90 & Ia & 0.07 & 5284 & 0.99 & 1.31 & 1.03 & 1.75 & S & Ia \\
13-SN-045 & SMC753.19.44N & 23:40:36.52 & -64:17:22.6 & 6515.81102 & 20.03 & 5277 & IIP & 0.70 & II & 0.10 & 5290 & - & - & - & - & - & IIP \\
13-SN-046 & SMC804.05.88N & 1:07:29.73 & -67:26:03.6 & 6514.80302 & 19.66 & 5277 & Ia & 0.90 & II & 0.07 & 5290 & - & - & - & - & S & IIP \\
13-SN-047 & SMC753.19.43N & 23:40:35.48 & -64:24:22.7 & 6515.81102 & 18.35 & 5277 & Ia & 1.00 & II & 0.06 & 5284 & 1.40 & 1.60 & 1.86 & 2.61 & - & IIP \\
13-SN-048 & SMC813.11.50N & 1:37:10.25 & -68:23:48.6 & 6519.75978 & 19.77 & 5292 & IIn & 0.53 & II & 0.06 & 5290 & 1.75 & 2.00 & 2.89 & 1.94 & S & IIP \\
13-SN-049 & MBR138.13.114N & 3:36:04.02 & -73:09:11.5 & 6519.86362 & 18.51 & 5292 & IIn & 0.67 & - & - & - & - & - & - & - & S & AGN-flare \\
13-SN-050 & SMC793.20.1N & 0:36:11.73 & -63:45:36.5 & 6520.71662 & 19.00 & 5292 & Ia & 0.32 & unkn & 0.13 & 5338 & 0.09 & 0.22 & 0.10 & 2.95 & S & GalaxyAGN \\
13-SN-051 & SMC791.12.33N & 0:32:19.55 & -66:25:44.3 & 6520.71233 & 18.73 & 5292 & Ia & 0.90 & Ia & 0.07 & 5335 & 1.80 & 2.38 & 1.81 & 1.98 & S & Ia \\
13-SN-052 & MBR118.20.216N & 2:39:53.91 & -73:07:05.3 & 6520.87379 & 19.48 & 5292 & Ia & 1.00 & - & - & - & 12.29 & - & 17.29 & 5.32 & S & Ia \\
13-SN-053 & MBR117.15.46N & 2:29:49.33 & -71:57:45.2 & 6520.87173 & 19.18 & 5321 & Ia & 1.00 & - & - & - & 0.89 & - & 1.77 & 2.98 & S & Ia \\
13-SN-054 & MBR122.04.597 & 2:51:33.57 & -73:00:49.3 & 6521.77577 & 19.83 & 5321 & Ia & 1.00 & - & - & - & 0.86 & - & 0.86 & 1.44 & S & Ia \\
13-SN-055 & SMC702.29.613 & 0:10:47.75 & -71:28:28.2 & 0.00000 & 19.68 & 5321 & Ia & 0.90 & - & - & - & - & - & - & - & S & Ia \\
13-SN-056 & LMC583.23.313 & 4:55:34.55 & -63:05:31.2 & 6522.93175 & 20.44 & 5321 & Ia & 0.94 & - & - & - & 0.13 & - & 0.16 & 1.54 & S & unknown \\
13-SN-057 & SMC793.02.42N & 0:37:02.41 & -64:29:50.0 & 6531.83447 & 19.40 & 5368 & IIP & 0.48 & Ia & 0.10 & 5345 & 1.25 & 2.28 & 1.28 & 1.45 & S & Ia \\
13-SN-058 & MBR200.11.30N & 2:51:13.59 & -63:59:33.8 & 6531.88532 & 17.61 & 5368 & CN & 0.70 & - & - & - & 7.11 & - & 8.98 & 4.01 & S & unknown-likelyIIn \\
13-SN-059 & MBR202.13.2N & 3:03:11.21 & -68:27:39.8 & 6526.81470 & 18.72 & 5368 & IIP & 0.45 & - & - & - & - & - & - & - & S & AGN \\
13-SN-060 & LMC602.04.31N & 5:48:11.18 & -61:42:15.4 & 6538.90401 & 18.63 & 5368 & Ia & 0.80 & - & - & - & - & - & - & - & S & Ia \\
13-SN-061 & LMC575.18.398 & 5:22:33.54 & -61:53:54.8 & 6533.89871 & 19.96 & 5368 & IIn & 0.53 & - & - & - & 1.23 & - & 1.36 & 5.78 & S & unknown \\
13-SN-062 & LMC601.20.1216 & 5:50:53.43 & -62:24:43.8 & 6538.90194 & 19.26 & 5368 & Ia & 0.36 & - & - & - & 0.29 & - & 0.30 & 1.14 & S & Ia \\
13-SN-063 & MBR203.14.488 & 3:01:54.01 & -67:07:49.6 & 6537.86792 & 19.63 & 5368 & IIP & 0.48 & - & - & - & - & - & - & - & - & IIP \\
13-SN-064 & MBR207.03.928 & 3:21:53.46 & -70:29:46.1 & 6537.87613 & 18.48 & 5368 & IIn & 0.67 & - & - & - & 2.99 & - & 3.00 & 3.27 & S & Ia \\
13-SN-065 & SMC824.07.741 & 0:56:40.86 & -80:16:33.3 & 6537.80629 & 19.68 & 5368 & IIP & 0.48 & - & - & - & 0.47 & - & 0.49 & 1.17 & S & unknown \\
13-SN-066 & LMC622.22.4N & 4:26:10.28 & -62:31:07.0 & 6543.92258 & 17.66 & 5397 & Ia & 0.80 & - & 0.017509 & NED & 0.22 & - & 0.24 & 3.91 & S & AGN \\
13-SN-067 & MBR135.28.39N & 3:39:46.50 & -75:47:33.6 & 6545.81138 & 19.83 & 5397 & IIn & 0.53 & - & - & - & 7.62 & - & 7.95 & 8.51 & S & unknown-likelyIIP \\
13-SN-068 & LMC517.03.1385N & 5:34:00.16 & -69:03:06.1 & 6547.87983 & 19.12 & 5397 & Ia & 0.36 & - & - & - & - & - & - & - & S & unknown-likelyIIn \\
13-SN-069 & SMC811.31.1104N & 1:15:50.70 & -63:17:17.3 & 6543.82157 & 19.97 & 5397 & IIn & 0.53 & - & - & - & - & - & - & - & S & IIP \\
13-SN-070 & SMC796.04.44N & 0:45:14.41 & -66:19:50.6 & 6545.75057 & 17.89 & 5397 & Ia & 1.00 & Ia & 0.043 & 5455 & - & - & - & - & - & Ia \\
13-NOVA-01 & MBR220.01.13N & 3:55:06.25 & -69:23:40.9 & 6523.83008 & 14.69 & - & CN & 1.00 & - & - & - & - & - & - & - & - & CN \\
13-SN-071 & LMC520.15.10585 & 5:26:09.30 & -65:08:13.4 & 6548.86404 & 19.13 & 5397 & IIP & 0.48 & - & - & - & 0.21 & - & 0.23 & 7.34 & S & unknown \\
13-SN-072 & MBR226.15.30N & 3:59:37.66 & -69:29:26.6 & 6546.86585 & 19.32 & 5445 & IIP & 1.00 & - & - & - & 0.31 & - & 0.33 & 7.10 & AGN & AGN-flare \\
13-SN-073 & MBR118.09.63N & 2:44:54.81 & -73:20:35.8 & 6564.71252 & 19.05 & 5445 & Ia & 0.90 & Ia & 0.091 & 5400 & 4.46 & 7.47 & 4.69 & 5.29 & S & Ia \\
13-SN-074 & MBR121.06.54N & 2:47:14.24 & -71:49:32.0 & 6563.74401 & 18.77 & 5445 & IIP & 0.48 & - & - & - & 2.56 & - & 5.71 & 6.67 & S & unknown \\
13-SN-075 & MBR153.07.76N & 1:58:01.88 & -68:45:47.9 & 6565.85535 & 18.95 & 5445 & Ia & 0.90 & Ia & 0.08 & 5440 & 4.67 & 6.97 & 6.02 & 4.55 & S & Ia \\
13-SN-076 & LMC555.32.3252N & 5:38:31.91 & -66:29:38.9 & 6561.85555 & 18.02 & 5445 & IIn & 0.97 & - & - & - & - & - & - & - & S & IIn \\
13-SN-077 & MBR179.21.47N & 2:08:49.55 & -67:22:50.5 & 6563.75965 & 20.58 & 5445 & IIP & 0.42 & - & - & - & - & - & - & - & - & unknown \\
13-SN-078 & SMC739.06.52N & 1:26:31.49 & -74:44:53.1 & 6565.66621 & 19.12 & 5445 & DN & 1.00 & - & - & - & 3.79 & - & 3.80 & 2.76 & AGN & Ia \\
13-SN-079 & SMC790.09.68N & 0:35:10.20 & -67:41:08.5 & 6565.68692 & 19.81 & 5445 & Ia & 1.00 & I & 0 & 5443 & - & - & - & - & S & unknown \\
13-SN-080 & SMC735.14.52N & 1:02:33.24 & -76:27:24.2 & 6565.65763 & 19.60 & 5445 & Ia & 0.99 & Ia & 0.103 & 5443 & 3.80 & 7.11 & 3.95 & 2.36 & S & Ia \\
13-SN-081 & LMC677.06.13N & 8:11:38.05 & -70:57:43.5 & 6562.88325 & 18.90 & 5455 & - & - & - & - & - & - & - & - & - & S & - \\
13-SN-082 & LMC549.01.26N & 4:38:39.51 & -66:13:58.2 & 6567.89397 & 18.55 & 5488 & IIn & 0.67 & - & 0.048867 & NED & - & - & - & - & S & AGN-flare \\
13-SN-083 & MBR110.12.51N & 2:05:48.86 & -75:42:26.2 & 6569.74423 & 20.37 & 5488 & IIP & 0.45 & - & - & - & - & - & - & - & S & unknown \\
13-SN-084 & MBR118.02.78N & 2:41:49.12 & -73:27:52.5 & 6569.76567 & 19.71 & 5488 & IIP & 1.00 & - & - & - & 7.08 & - & 7.12 & 5.03 & S & Ia \\
13-SN-085 & MBR187.22.207N & 2:18:14.52 & -64:56:59.6 & 6568.83820 & 19.25 & 5488 & Ia & 1.00 & - & - & - & 0.57 & - & 0.66 & 3.16 & S & Ia \\
13-SN-086 & SMC797.20.360N & 0:46:36.36 & -64:31:06.7 & 6569.69769 & 20.35 & 5488 & Ia & 0.97 & - & - & - & 5.33 & - & 5.42 & 7.01 & S & unknown \\
13-SN-087 & LMC534.12.38N & 4:55:44.52 & -65:40:27.8 & 6571.87145 & 19.71 & 5488 & Ia & 0.36 & - & - & - & 4.62 & - & 5.52 & 1.70 & Ell & unknown \\
\hline
\caption{Properties of all transients found by the OGLE-IV Transients Detection System in seasons 2012-2014. See main text for details.}
\end{longtable}
\end{tiny}
\end{landscape}
\begin{landscape}
\begin{tiny}
\begin{longtable}{l l c c c c c c c c c c c c c c c c}
\hline
ID & DBID & RA$_\mathrm{J2000.0}$ & DEC$_\mathrm{J2000.0}$ & Discovery & Disc. & ATEL & phot. & prob. & spec. & z & ATEL  & Offset & Offset & Offset$^*$ & $R_\mathrm{Ser}$ & WISE & comment \\
 &  &  &  & HJD-2450000 & $I$ mag & \# & class & & type &  & spec.\#  & [arcsec] & [kpc] & [arcsec] & [arcsec] & & \\
\hline
13-SN-088 & LMC548.07.41N & 4:27:44.75 & -67:24:22.4 & 6573.79563 & 19.59 & 5488 & IIP & 0.48 & AGN & 0.168 & 5459 & - & - & - & - & AGN & AGN \\
13-SN-089 & LMC599.26.1289N & 5:42:37.14 & -60:09:12.2 & 6572.84328 & 19.47 & 5488 & IIP & 0.48 & - & - & - & 6.36 & - & 6.71 & 5.13 & S & unknown \\
13-SN-090 & MBR229.10.1N & 4:08:38.82 & -65:45:59.2 & 6571.75868 & 18.64 & 5488 & Ia & 0.80 & AGN & 0.125 & 5537 & 0.10 & 0.23 & 0.11 & 5.45 & AGN & AGN \\
13-SN-091 & LMC596.26.1733N & 5:44:44.23 & -63:55:51.1 & 6575.87438 & 18.51 & 5488 & Ia & 0.80 & Ic & 0.08 & 5510 & 6.32 & 9.42 & 7.67 & 9.22 & Ell & Ic \\
13-SN-092 & SMC742.20.305N & 23:17:21.63 & -77:53:55.5 & 6572.62264 & 20.13 & 5488 & Ia & 1.00 & - & - & - & - & - & - & - & - & Ia \\
13-SN-093 & LMC503.19.415N & 5:23:09.14 & -69:03:46.2 & 6571.84517 & 19.81 & 5488 & Ia & 0.36 & - & - & - & - & - & - & - & S & Ia \\
13-SN-094 & LMC644.05.41N & 5:34:48.25 & -76:38:39.8 & 6573.80699 & 19.52 & 5488 & Ia & 1.00 & - & - & - & 0.47 & - & 0.52 & 1.10 & S & unknown \\
13-NOVA-02 & LMC649.28.33N & 5:57:58.35 & -74:54:08.9 & 6771.49064 & 11.55 & - & CN & 0.90 & - & - & - & - & - & - & - & - & CN \\
13-SN-095 & LMC657.18.66N & 6:43:01.30 & -76:25:11.2 & 6584.87665 & 18.77 & 5497 & DN & 1.00 & - & - & - & - & - & - & - & S & DN \\
13-SN-096 & LMC599.20.159N & 5:39:38.80 & -60:39:32.1 & 6584.78940 & 19.60 & 5497 & Ia & 0.90 & Ia & 0.11 & 5510 & 5.89 & 11.68 & 8.01 & 4.60 & S & Ia \\
13-SN-097 & LMC548.19.44N & 4:34:33.92 & -66:46:42.7 & 6584.83860 & 19.77 & 5497 & Ia & 0.90 & - & - & - & 1.87 & - & 2.54 & 14.39 & S & unknown \\
13-SN-098 & LMC712.04.7N & 5:26:36.40 & -49:28:03.9 & 6583.86939 & 18.70 & 5497 & Ia & 0.90 & Ia & 0.06 & 5504 & 1.15 & 1.32 & 1.50 & 3.99 & S & Ia \\
13-SN-099 & LMC676.12.15N & 8:02:27.32 & -72:25:26.6 & 6585.87341 & 17.71 & 5497 & Ia & 1.00 & Ia & 0.028 & 550 & 3.71 & 2.06 & 4.35 & 7.84 & S & Ia \\
13-SN-100 & MBR114.24.1106 & 2:16:18.56 & -74:45:23.7 & 6586.67071 & 18.89 & 5497 & IIP & 0.48 & II-pec & 0.09 & 5617 & 1.05 & 1.74 & 1.11 & 2.57 & S & unknown-peculiar \\
13-SN-101 & LMC579.20.639 & 5:10:10.57 & -62:31:29.2 & 6584.76180 & 19.84 & 5497 & IIP & 0.70 & - & - & - & 2.29 & - & 2.90 & 3.78 & S & unknown \\
13-SN-102 & LMC644.24.2119 & 5:31:12.75 & -75:50:49.6 & 6584.84950 & 19.11 & 5497 & IIP & 0.70 & - & - & - & 0.00 & - & 0.01 & 0.44 & - & unknown-likely DN \\
13-SN-103 & MBR121.06.808 & 2:46:12.84 & -71:50:45.0 & 6585.75997 & 18.71 & 5497 & IIP & 0.48 & - & - & - & 2.10 & - & 2.48 & 2.53 & S & unknown/I/IIb \\
13-SN-104 & MBR171.22.812 & 3:22:00.48 & -76:34:11.6 & 6586.69717 & 19.15 & 5497 & Ia & 1.00 & - & - & - & 0.19 & - & 0.21 & 2.24 & S & Ia \\
13-SN-105 & MBR213.27.331 & 3:38:44.20 & -69:06:01.4 & 6585.77689 & 19.80 & 5497 & Ia & 0.90 & - & 0.051479 & NED & 14.89 & - & 14.92 & 9.04 & S & unknown \\
13-SN-106 & SMC770.06.363 & 23:53:06.84 & -78:33:19.5 & 6583.58641 & 19.30 & 5497 & Ia & 0.98 & - & - & - & 1.20 & - & 1.44 & 2.20 & S & Ia \\
13-SN-107 & SMC787.23.110N & 0:17:38.87 & -65:39:34.8 & 6595.60489 & 19.84 & 5535 & Ia & 1.00 & - & - & - & - & - & - & - & - & Ia \\
13-SN-108 & SMC784.27.66N & 0:14:10.58 & -63:19:31.2 & 6595.59871 & 18.90 & 5535 & IIn & 0.42 & - & - & - & 0.46 & - & 0.46 & 2.26 & S & unknown \\
13-SN-109 & MBR148.22.532N & 1:46:09.34 & -67:28:00.1 & 6595.66504 & 19.22 & 5535 & Ia & 0.99 & Ia & 0.088 & 5537 & 3.70 & 6.02 & 3.71 & 12.70 & S & Ia \\
13-SN-110 & SMC801.19.64N & 0:59:40.13 & -65:04:57.2 & 6600.60124 & 18.84 & 5581 & Ia & 0.90 & - & - & - & (2.1) & - & - & - & - & Ia \\
13-SN-111 & SMC765.15.1578 & 23:42:18.27 & -67:39:38.8 & 6604.52732 & 19.78 & 5581 & Ia & 1.00 & - & - & - & 1.04 & - & 2.34 & 3.52 & S & Ia \\
13-SN-112 & LMC519.30.237N & 5:29:47.48 & -65:44:13.7 & 6608.84498 & 19.00 & 5581 & IIP & 0.48 & - & - & - & - & - & - & - & S & IIn \\
13-SN-113 & LMC550.07.4184 & 5:44:32.80 & -73:32:08.1 & 6607.80106 & 19.96 & 5581 & IIP & 0.70 & - & - & - & 0.60 & - & 1.10 & 1.31 & S & unknown \\
13-SN-114 & SMC700.19.63N & 0:19:35.61 & -69:24:12.5 & 6609.51537 & 19.45 & 5581 & Ia & 0.99 & - & - & - & 2.29 & - & 2.65 & 3.51 & S & Ia \\
13-SN-115 & MBR126.32.2978N & 3:00:56.46 & -70:16:47.9 & 6610.68031 & 19.78 & 5581 & IIP & 0.53 & - & - & - & - & - & - & - & S & unknown \\
13-SN-116 & MBR167.02.21N & 3:05:37.43 & -79:16:57.8 & 6610.63029 & 19.07 & 5581 & Ia & 0.32 & - & - & - & 2.01 & - & 2.02 & 3.82 & Ell & Ia \\
13-SN-117 & LMC640.28.6330 & 5:21:53.08 & -74:55:42.7 & 6614.83575 & 18.12 & 5608 & IIn & 0.67 & - & - & - & 1.16 & - & 1.43 & 2.42 & S & Ia \\
13-SN-118 & LMC505.24.64N & 5:14:47.50 & -66:50:29.1 & 6617.72346 & 19.01 & 5608 & Ia & 1.00 & Ia & 0.07 & 5595 & 6.18 & 8.16 & 6.81 & 2.75 & S & Ia \\
13-SN-119 & LMC566.28.27N & 6:19:45.55 & -71:20:19.4 & 6617.73597 & 19.98 & 5608 & IIP & 0.42 & - & - & - & - & - & - & - & Ell & DN \\
13-SN-120 & LMC556.29.2646 & 5:41:49.26 & -65:13:05.4 & 6618.81650 & 18.95 & 5608 & Ia & 0.90 & Ia & 0.07 & 5595 & - & - & - & - & S & Ia \\
13-SN-121 & LMC616.26.2478 & 4:22:42.87 & -68:16:34.2 & 6618.72813 & 21.17 & 5608 & IIP & 0.42 & - & - & - & 9.21 & - & 9.25 & 7.24 & S & unknown \\
13-SN-122 & LMC620.21.311 & 4:14:53.34 & -63:58:35.1 & 6618.73757 & 21.16 & 5608 & IIP & 0.70 & - & - & - & 0.04 & - & 0.04 & 1.22 & S & unknown \\
13-SN-123 & LMC604.05.69N & 5:58:30.38 & -63:33:38.3 & 6616.78872 & 18.36 & 5608 & IIn & 0.67 & Ia & 0.08 & 5602 & 0.16 & 0.23 & 0.16 & 1.99 & Ell & Ia \\
13-SN-124 & MBR201.19.51N & 3:08:52.84 & -69:25:09.7 & 6622.68027 & 20.22 & 5608 & Ia & 0.90 & Ia & 0.13 & 5615 & (29.6) & - & - & - & - & Ia \\
13-SN-125 & LMC621.26.1477N & 4:30:02.84 & -63:11:29.9 & 6622.74714 & 19.94 & 5608 & Ia & 0.90 & - & - & - & 12.53 & - & 12.53 & 2.75 & S & Ia \\
13-SN-126 & LMC620.18.111N & 4:19:47.17 & -63:43:21.6 & 6622.74502 & 19.15 & 5608 & Ia & 1.00 & Ia & 0.06 & 5615 & - & - & - & - & - & Ia \\
13-SN-127 & MBR203.11.294 & 3:05:57.62 & -67:08:18.5 & 6622.68470 & 20.74 & 5608 & Ia & 0.94 & - & - & - & 1.02 & - & 1.02 & 1.41 & S & Ia \\
13-SN-128 & LMC636.05.3425 & 4:59:42.75 & -75:18:04.8 & 6623.79520 & 19.24 & 5608 & IIP & 0.48 & - & - & - & 2.09 & - & 2.45 & 3.95 & S & unknown/IIP \\
13-SN-129 & SMC787.06.77N & 0:16:41.97 & -66:16:07.4 & 6625.56369 & 19.45 & 5663 & Ia & 0.99 & Ia & 0.08 & 5622 & 11.54 & 17.20 & 11.81 & 4.92 & S & Ia \\
13-SN-130 & LMC615.28.29N & 6:36:46.99 & -68:52:22.4 & 6627.85385 & 19.65 & 5663 & IIP & 0.48 & Ia-pec & 0.09 & 5620 & - & - & - & - & S & Ia-pec \\
13-SN-131 & MBR149.05.146N & 1:43:36.96 & -66:50:39.7 & 6629.62253 & 19.48 & 5663 & Ia & 0.90 & - & - & - & 4.03 & 4.62 & 5.82 & 4.82 & S & unknown \\
13-SN-132 & LMC612.26.1546N & 6:34:30.72 & -70:40:00.3 & 6635.81133 & 19.69 & 5663 & Ia & 1.00 & - & - & - & 5.53 & - & 5.54 & 3.45 & S & Ia \\
\hline
\caption{Properties of all transients found by the OGLE-IV Transients Detection System in seasons 2012-2014. See main text for details.}
\end{longtable}
\end{tiny}
\end{landscape}
\begin{landscape}
\begin{tiny}
\begin{longtable}{l l c c c c c c c c c c c c c c c c}
\hline
ID & DBID & RA$_\mathrm{J2000.0}$ & DEC$_\mathrm{J2000.0}$ & Discovery & Disc. & ATEL & phot. & prob. & spec. & z & ATEL  & Offset & Offset & Offset$^*$ & $R_\mathrm{Ser}$ & WISE & comment \\
 &  &  &  & HJD-2450000 & $I$ mag & \# & class & & type &  & spec.\#  & [arcsec] & [kpc] & [arcsec] & [arcsec] & & \\
\hline
13-SN-133 & LMC560.22.309N & 5:57:43.83 & -69:50:06.2 & 6636.81928 & 19.43 & 5663 & IIP & 0.90 & - & - & - & - & - & - & - & S & IIP \\
13-SN-134 & LMC565.24.54N & 6:13:58.04 & -72:58:24.9 & 6636.81237 & 18.85 & 5663 & Ia & 0.90 & Ic-pec & 0.039 & 5696 & - & - & - & - & S & Ic-pec \\
13-SN-135 & MBR168.20.211N & 3:06:48.20 & -77:24:28.4 & 6636.60643 & 19.22 & 5663 & IIP & 0.92 & II & 0.057 & 5718 & 3.46 & 3.77 & 3.52 & 3.08 & S & IIP \\
13-SN-136 & LMC582.05.55N & 4:55:46.85 & -64:51:20.4 & 6639.75154 & 19.33 & 5663 & Ia & 1.00 & Ia & 0.080 & 5696 & 4.42 & 6.60 & 10.70 & 8.99 & S & Ia \\
13-SN-137 & MBR108.05.107N & 2:03:34.72 & -73:36:22.7 & 6639.60394 & 19.14 & 5663 & IIP & 0.48 & IIn & 0.075 & 5686 & 5.85 & 8.23 & 6.73 & 6.49 & S & IIn \\
13-SN-138 & SMC729.32.3036N & 1:19:02.87 & -68:13:14.7 & 6638.56148 & 18.59 & 5663 & IIn & 0.94 & IIn & 0.115 & 5683 & - & - & - & - & S & IIn \\
13-SN-139 & LMC677.15.55N & 8:09:25.36 & -70:37:59.3 & 6640.84276 & 20.07 & 5663 & Ia & 1.00 & - & - & - & 3.38 & - & 3.39 & 2.26 & S & Ia \\
13-SN-140 & LMC636.21.5988 & 5:02:52.42 & -74:40:11.3 & 6639.71997 & 20.20 & 5663 & IIP & 0.70 & - & - & - & 0.01 & - & 0.01 & 0.43 & - & Ia \\
13-NOVA-03 & MBR115.04.1856 & 2:26:43.16 & -76:34:37.8 & 6637.60535 & 17.15 & - & CN & 0.80 & - & - & - & - & - & - & - & - & CN \\
13-SN-141 & MBR171.15.67 & 3:16:00.60 & -77:05:38.5 & 6640.62861 & 18.96 & 5663 & Ia & 0.90 & Ia & 0.05 & 5696 & 0.09 & 0.08 & 0.23 & 5.05 & S & Ia \\
13-SN-142 & MBR162.24.72N & 2:10:37.64 & -78:01:50.2 & 6640.58723 & 19.34 & 5697 & Ia & 0.90 & - & - & - & 3.87 & - & 4.58 & 4.62 & S & Ia \\
13-SN-143 & LMC657.11.2388 & 6:37:30.46 & -76:45:00.1 & 6645.76906 & 20.56 & 5697 & Ia & 0.88 & - & - & - & 1.60 & - & 1.63 & 1.80 & S & Ia \\
13-SN-144 & LMC511.15.104N & 5:00:57.18 & -67:44:25.7 & 6647.74442 & 18.76 & 5697 & Ia & 1.00 & II & 0.04 & 5696 & - & - & - & - & S & IIP/IIL \\
13-SN-145 & LMC547.20.284N & 4:30:38.05 & -67:54:31.2 & 6647.70245 & 19.80 & 5697 & Ia & 0.80 & - & - & - & 9.17 & - & 9.22 & 3.50 & S & unknown \\
13-SN-146 & LMC548.11.58N & 4:33:00.74 & -67:05:21.5 & 6647.70467 & 18.69 & 5697 & Ia & 1.00 & - & - & - & - & - & - & - & S & Ia \\
13-SN-147 & LMC563.07.67N & 5:48:57.86 & -66:40:15.7 & 6649.78517 & 19.06 & 5697 & IIP & 0.48 & Ia-pec & 0.099 & 5689 & 0.08 & 0.14 & 0.09 & 6.57 & S & Ia-pec \\
13-SN-148 & LMC658.10.39N & 6:38:06.94 & -75:43:36.6 & 6652.70473 & 19.78 & 5697 & Ia & 1.00 & Ia & 0.043 & 5701 & - & - & - & - & S & Ia \\
13-SN-149 & LMC708.12.33N & 5:27:36.13 & -54:04:25.2 & 6653.59944 & 20.02 & 5709 & Ia & 1.00 & II & 0.079 & 5720 & 6.34 & 9.34 & 6.38 & 4.58 & S & IIn \\
13-SN-150 & LMC504.30.333N & 5:17:28.40 & -67:33:29.7 & 6654.76446 & 20.15 & 5709 & Ia & 0.90 & - & - & - & - & - & - & - & Ell & Ia \\
13-SN-151 & LMC517.28.304N & 5:34:24.69 & -68:10:07.8 & 6654.77073 & 20.08 & 5709 & IIP & 0.92 & - & - & - & - & - & - & - & S & Ia \\
13-SN-152 & MBR134.03.729N & 3:33:38.89 & -75:21:16.8 & 6656.64804 & 19.20 & 5709 & Ia & 0.90 & - & - & - & - & - & - & - & S & IIP \\
13-SN-153 & SMC792.28.83N & 0:34:37.69 & -64:42:10.1 & 6656.53603 & 18.75 & 5709 & Ia & 1.00 & - & - & - & 2.43 & - & 3.61 & 3.21 & S & Ia \\
13-SN-154 & LMC654.04.36N & 6:14:28.06 & -75:23:53.3 & 6657.67448 & 20.49 & 5709 & IIn & 0.51 & - & - & - & 0.13 & - & 0.13 & 1.11 & S & unknown \\
13-SN-155 & LMC669.10.39N & 7:35:27.88 & -71:15:07.1 & 6657.84450 & 20.45 & 5709 & Ia & 0.71 & - & - & - & 2.38 & - & 2.40 & 2.06 & - & IIP or IIn \\
13-SN-156 & LMC583.15.1914 & 4:54:08.66 & -63:15:37.1 & 6657.80890 & 20.31 & 5709 & Ia & 0.90 & Ia & 0.14 & 5716 & 0.03 & 0.08 & 0.07 & 1.73 & S & Ia \\
14-SN-001 & LMC664.05.78N & 7:02:35.58 & -72:48:15.9 & 6659.75454 & 19.07 & 5817 & Ia & 0.32 & CV & 0 & 5718 & - & - & - & - & S & DN \\
14-SN-002 & MBR231.04.42N & 4:04:31.71 & -63:50:15.6 & 6660.72357 & 19.53 & 5817 & Ia & 0.90 & Ia & 0.10 & 5748 & 2.75 & 5.01 & 3.34 & 4.90 & S & Ia \\
14-SN-003 & MBR108.19.69N & 2:09:59.06 & -73:02:11.8 & 6664.62577 & 19.10 & 5817 & IIP & 0.48 & - & - & - & 3.23 & - & 3.73 & 2.34 & S & unknown \\
14-SN-004 & LMC549.18.139N & 4:38:36.94 & -65:31:29.8 & 6668.76158 & 18.79 & 5817 & IIP & 0.92 & IIP & 0.03 & 5912 & 6.08 & 3.60 & 6.12 & 6.28 & S & IIP \\
14-SN-005 & LMC665.32.967 & 7:12:06.72 & -70:07:13.2 & 6682.73417 & 21.60 & 5817 & Ia & 0.97 & - & - & - & 0.21 & - & 0.25 & 6.82 & S & unknown \\
14-SN-006 & LMC671.21.46N & 7:45:41.68 & -70:18:58.4 & 6682.74720 & 19.87 & 5817 & IIP & 1.00 & - & - & - & 0.07 & - & 0.10 & 1.84 & S & IIP \\
14-SN-007 & MBR117.07.95N & 2:29:31.88 & -72:14:03.7 & 6690.54118 & 19.04 & 5875 & Ia & 0.99 & - & - & - & (1.8) & - & - & - & - & unknown \\
14-SN-008 & LMC578.23.56N & 5:05:52.96 & -63:44:54.4 & 6693.70282 & 19.52 & 5875 & Ia & 1.00 & - & - & - & 10.59 & - & 10.63 & 3.75 & AGN & unknown \\
14-SN-009 & LMC708.18.143N & 5:29:53.65 & -53:53:16.6 & 6694.74193 & 19.57 & 5875 & Ia & 1.00 & II & 0.056 & 5864 & 1.46 & 1.57 & 1.55 & 6.28 & S & unknown \\
14-SN-010 & LMC614.07.2860 & 6:25:23.80 & -68:34:20.3 & 6694.77028 & 18.95 & 5875 & Ia & 1.00 & Ia & 0.081 & 5864 & 0.08 & 0.12 & 0.08 & 0.99 & - & Ia \\
14-SN-011 & LMC659.30.53N & 6:46:58.87 & -69:57:24.8 & 6695.72994 & 19.89 & 5875 & IIP & 0.70 & IIn & 0.083 & 5908 & 0.81 & 1.25 & 1.34 & 1.72 & S & IIn \\
14-SN-012 & LMC668.31.940N & 7:27:43.39 & -69:20:35.0 & 6695.74927 & 20.24 & 5875 & IIn & 0.51 & - & - & - & - & - & - & - & - & unknown/IIP \\
14-SN-013 & LMC673.32.2006N & 7:40:35.37 & -72:28:17.1 & 6695.76024 & 20.17 & 5875 & Ia & 1.00 & - & - & - & - & - & - & - & - & Ia \\
14-SN-014 & LMC522.21.92N & 4:27:23.75 & -74:42:11.1 & 6698.66286 & 18.91 & 5875 & IIP & 0.90 & Ib & 0.05 & 5891 & 6.17 & 5.96 & 6.43 & 5.60 & S & Ib-unknown \\
14-SN-015 & LMC610.19.3006 & 6:25:23.80 & -68:34:20.3 & 6700.70514 & 18.98 & 5916 & Ia & 1.00 & - & - & - & - & - & - & - & - & Ia \\
14-SN-016 & LMC673.31.1107N & 7:42:01.43 & -72:29:41.4 & 6706.74447 & 19.41 & 5916 & Ia & 1.00 & Ia & 0.07 & 5908 & - & - & - & - & S & Ia \\
14-SN-017 & MBR218.07.74N & 3:30:32.38 & -63:43:16.1 & 6709.61149 & 19.07 & 5916 & Ia & 1.00 & - & - & - & - & - & - & - & AGN & Ia \\
14-SN-018 & LMC608.19.61N & 6:18:20.35 & -66:01:08.0 & 6710.70753 & 18.37 & 5916 & IIn & 0.67 & II & 0.03 & 5915 & - & - & - & - & S & IIP \\
14-SN-019 & LMC609.07.75N & 6:13:48.04 & -67:55:15.0 & 6710.70958 & 17.41 & 5916 & Ia & 1.00 & Ia & 0.04 & 5915 & 0.52 & 0.41 & 0.67 & 7.41 & Ell & Ia \\
14-SN-020 & LMC577.21.2359 & 6:24:30.22 & -73:58:10.6 & 6710.63295 & 20.27 & 5916 & IIP & 0.70 & II & 0.076 & 5934 & 0.04 & 0.06 & 0.09 & 0.82 & S & IIP \\
14-SN-021 & LMC555.18.230N & 5:48:23.49 & -66:47:29.7 & 6712.66078 & 17.92 & 5916 & Ia & 1.00 & Ia & 0.039 & 5919 & 2.80 & 2.14 & 3.20 & 2.21 & S & Ia \\
\hline
\caption{Properties of all transients found by the OGLE-IV Transients Detection System in seasons 2012-2014. See main text for details.}
\end{longtable}
\end{tiny}
\end{landscape}
\begin{landscape}
\begin{tiny}
\begin{longtable}{l l c c c c c c c c c c c c c c c c}
\hline
ID & DBID & RA$_\mathrm{J2000.0}$ & DEC$_\mathrm{J2000.0}$ & Discovery & Disc. & ATEL & phot. & prob. & spec. & z & ATEL  & Offset & Offset & Offset$^*$ & $R_\mathrm{Ser}$ & WISE & comment \\
 &  &  &  & HJD-2450000 & $I$ mag & \# & class & & type &  & spec.\#  & [arcsec] & [kpc] & [arcsec] & [arcsec] & & \\
\hline
14-SN-022 & MBR145.13.195N & 4:03:27.76 & -72:11:27.0 & 6728.57155 & 17.10 & - & IIn & 0.67 & IIn & 0.024 & 5934 & 7.13 & 3.40 & 7.38 & 4.83 & S & IIn \\
14-SN-023 & LMC680.22.414N & 8:26:16.27 & -69:52:21.5 & 6715.73155 & 18.98 & - & DN & 1.00 & CV & 0 & 5937 & - & - & - & - & - & DN \\
14-SN-024 & LMC647.02.50N & 6:01:15.38 & -78:18:21.7 & 6719.64051 & 19.69 & - & IIP & 0.48 & Ia & 0.1 & 5938 & - & - & - & - & S & Ia \\
14-SN-025 & LMC689.24.78N & 5:07:07.33 & -54:57:24.6 & 6721.58127 & 20.34 & - & Ia & 1.00 & - & - & - & 1.55 & - & 2.10 & 2.74 & S & Ia \\
14-SN-026 & LMC697.18.174N & 5:22:13.47 & -56:46:47.6 & 6721.55699 & 20.10 & - & IIP & 0.70 & - & - & - & 2.13 & - & 2.37 & 5.15 & S & IIP \\
14-SN-027 & MBR164.04.58N & 2:41:49.91 & -78:24:41.8 & 6728.51211 & 18.62 & - & IIn & 0.41 & - & - & - & 0.94 & - & 0.94 & 2.41 & S & unknown \\
14-SN-028 & LMC635.01.33N & 5:09:42.76 & -76:41:29.6 & 6734.55975 & 20.17 & - & - & - & - & - & - & 0.40 & - & 0.46 & 2.58 & Ell & - \\
14-NOVA-01 & BLG643.22.8N & 17:55:20.27 & -23:23:54.5 & 6735.89972 & 16.29 & - & - & - & - & - & - & - & - & - & - & - & - \\
\hline
\caption{Properties of all transients found by the OGLE-IV Transients Detection System in seasons 2012-2014. See main text for details.}
\end{longtable}
\end{tiny}
\end{landscape}

\section{Classification of transients}
During the real-time detection in years 2012-2014 the transients were classified only by a human observer and were split into two classes of candidates: supernovae or novae, typically relying on contextual information (primarily by checking if there was a galaxy-like object nearby) and the observed amplitudes.  
More robust and detailed classification can be performed either {\it via} spectroscopy while the transient is still on-going, or after the event is over and the full light evolution can be used for distinguishing transient classes. 
Below we describe those two channels.

\subsection{Spectroscopy}
\label{sec:specclass}
OGLE-IV transients appeared on the web-site after being found in the last-night's data (typically sooner than 12h after the last observation) and were immediately available for the astronomical community for the follow-up. 
Spectroscopic follow-up was carried out mostly by the Public ESO Survey PESSTO\footnote{http://www.pessto.org}(Valenti \etal 2012, Fraser \etal 2013, Smartt \etal 2013), using the New Technology Telescope (NTT) at La Silla Observatory, located about 20 km from Las Campanas. 
On a few occasions, however, OGLE transients were classified by PESSTO during the same night they were discovered, allowing for early confirmation of their nature and providing useful observational data for studying the early stages of supernova evolution.
Another two sources of spectroscopic classification of the OGLE transients were the Carnegie Supernova Project (CSP, Hamuy \etal 2006) using the 6.5m Magellan Telescope and 100 inch Du Pont telescope in Las Campanas Observatory, and observers from the Australian National University (ANU) using the Wide-Field Spectrograph (WiFeS) at ANU 2.3m telescope in Siding Spring Observatory.
The spectra of the OGLE supernovae classified by PESSTO are available {\it via} WiseRep repository (Yaron \etal 2013), whereas all the remaining spectra should be available from the respective groups.
Spectral typing, redshifts and reference to the relevant classification telegram are given in Table \tabmain.

Out of 238 transients reported by OGLE-IV real-time pipeline since 11th October 2012 until end of the 2013/2014 season (March 2014), 87 were classified spectroscopically and 83 turned out to be supernovae. 
Among those there were 47 Type Ia, 27 Type II (including 10 Type IIn and 2 Type IIp), as well as single examples of Ib, Ibn, Ic and peculiar Ia and Ic. 
The redshift range was from 0.014 (OGLE12-009) to 0.14 (OGLE13-043) for supernovae and up to 0.168 for AGNs (OGLE13-088).
Table \tabmain~contains information on the spectral classification and derived redshift.
Examples of different types of supernovae and CVs are shown in Figs. \figLCSNe~and \figLCNDN.

\begin{figure}
\centering
\includegraphics[width=1\columnwidth]{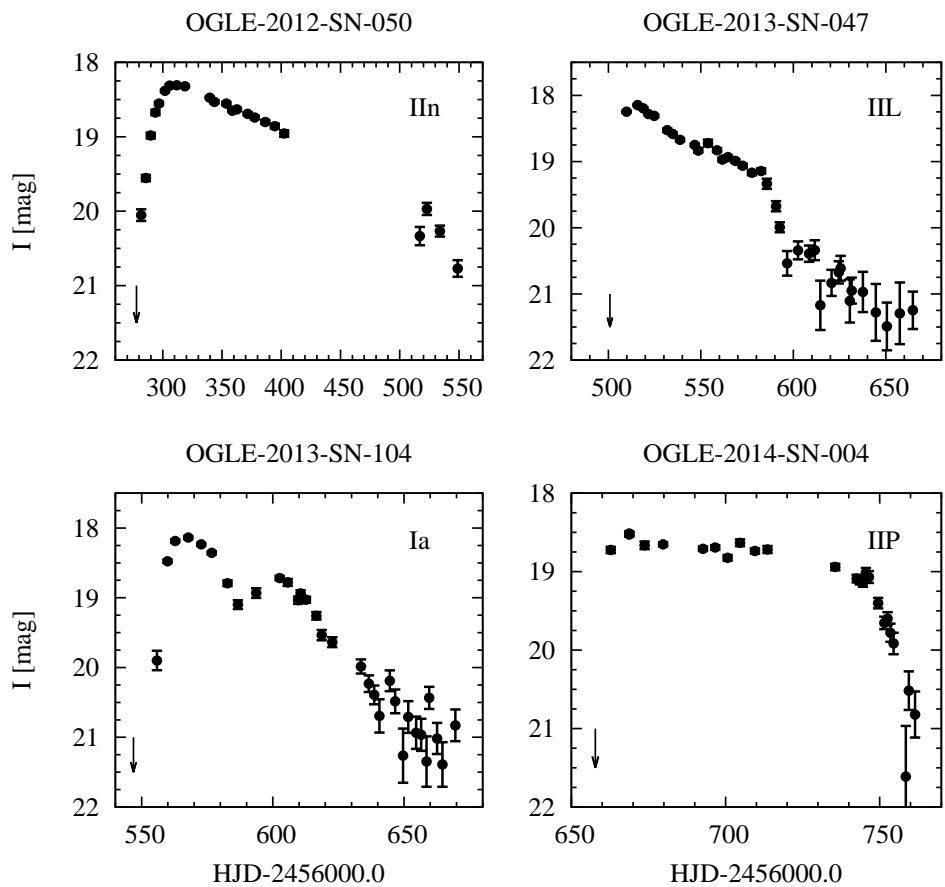}
\caption{Examples of various spectroscopically determined supernova types among transients detected by the OGLE-IV Transients Detection System. 
}
\label{fig:LC2}
\end{figure}

\begin{figure}
\centering
\includegraphics[width=1\columnwidth]{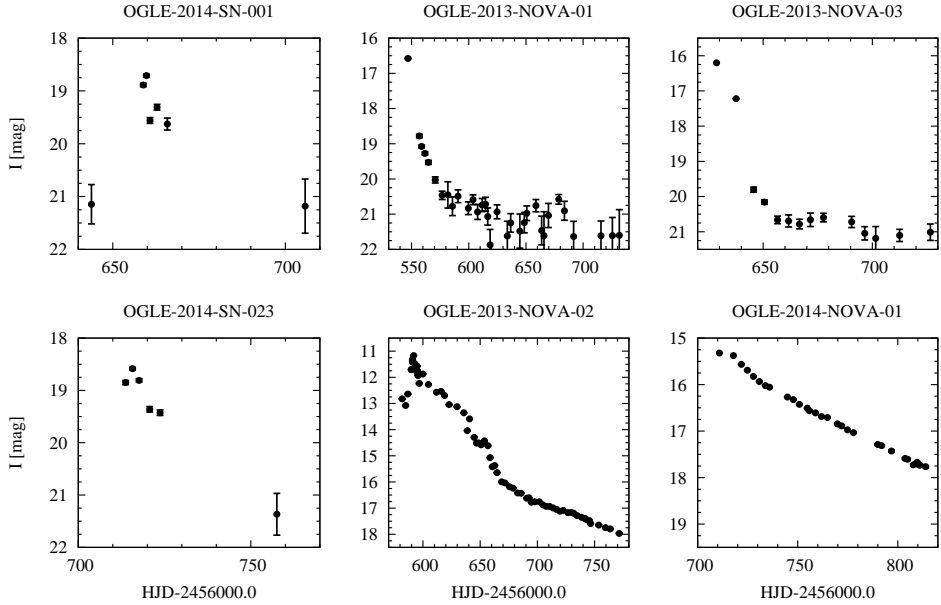}
\caption{Examples of Novae and Dwarf Novae detected by the OGLE-IV Transients Detection System.}
\label{fig:LC3}
\end{figure}

\subsection{Photometry and Machine Learning Classification}
\label{sec:photclass}
Thanks to its relatively high cadence, OGLE-IV provides well sampled light curves of supernovae and other transients in the $I$-band and more sparse light curves in $V$-band.
Therefore, the data can reveal characteristic features allowing us to distinguish between different supernova classes and cataclysmic variables. 
In particular, obtaining an early preliminary classification of a transient candidate could be helpful in allocating limited spectroscopic follow-up resources.
Within the OGLE-IV Transients Detection System pipeline we typically detect transients near or just after their maxima,  and there are at least a couple of data points available, as well as the time of the last non-detection.
Here we present and test the performance of an automated light curve classifier, which uses such incomplete early data as input.
This classifier will be implemented in the processing and detection chain in subsequent observing seasons. 

In order to build a training set for the classifier we used both spectral classification and (somewhat subjective) visual classification of the full light curves and separated our findings into five major classes of transients: supernovae Type Ia (Ia), core-collapse supernovae of Type IIn and IIp, dwarf novae (DNe) and classical novae (CNe).
The visual inspection was carried out by multiple experienced observers and relied on identifying crucial features of each of the classified types (\eg second maximum in SNe Type Ia), as well as contextual information (\eg presence of a galaxy near the transient and galaxy type). 
The results of the visual classification are listed in Table \tabmain~under {\it comment} column. 
The visual classification was attempted on the entire set of transients, however, still many objects remained classified as {\it unknown}.
Among those visually classified, we selected about a dozen from each class to construct the training set. 
We would like to emphasize here the advantages of the OGLE data collected in the $I$-band, which allow to maintain high purity of the photometric classification thanks to clear second maximum present in SNe Type Ia and well sampled light curves.

Because there were only a couple of Dwarf Novae found by the Transient Detection System (usually discarded during the detection process and not reported on the web-site), for training we used a set of synthetic light curves, generated from a linear rise and exponential decline model of a DN outburst based on several dozens of DNe found in the OGLE-III data by Skowron \etal (2009).
Several light curves were excluded from the training set because they had no pre-maximum data or the last non-detection date was unknown (for the very first transients).
The training set for the automated classifier comprised of 63 objects of Type Ia and II SNe, CNe and about a hundred of simulated DNe, adjusted to the OGLE-IV sampling.
Each light curve in the training set was trimmed at its maximum to mimic its typical appearance at the discovery epoch.
For such data we computed the following set of features:
\begin{itemize}
\item{slope1}- slope in mag/d before the max,
\item{mag1}- the maximum observed brightness,
\item{time2max}- time to reach the maximum from the last non-detection,
\item{rise}- difference in mag between the max and the detection level.
\end{itemize}

We classified the $I$-band photometry using a Random Forest classifier (Breiman \etal 2001) as implemented in the Weka package \footnote{{\it http://www.cs.waikato.ac.nz/ml/weka/} Weka, version 3.6.5, developed at the University of Waikato in New Zealand}.
Random Forest (RF) takes as an input a set of values, which can represent any feature of the light curve.
We trained the RF model on the training set and then classified all light curves with trained model.
The highest ranked (winner) class and its probability are provided in Table \tabmain in the {\it phot.class} column, along with its probability ($prob.$ column).
Note, the light curve classification was performed in the observer frame, however, this had a negligible impact on the result as the redshifts of most of our extragalactic transients were ranging from z=0.05--0.15. 

There are in total 238 transients in our table, however, if we  exclude all transients with uncertain visual classification or with not enough data points before the maximum brightness, we are left with 196 objects.
Further on, if we exclude objects classified outside of our five classes (\eg AGNs), we are left with 143 classifiable transients.
Among those, 120 (84\%) were assigned a class in agreement with the spectral and visual classifications. 
This is a very promising result, especially given the fact, that in some cases there was just 1 data point between a non-detection and the maximum brightness.
Overall, the performance of the classifier is good enough to provide not only distinction between CVs and SNe, but also between thermonuclear and core-collapse supernovae and their major subtypes.
The sample of CNe used for training of the classifier should, however, be extended in future, with new detections, but also possibly with the OGLE-IV data from the Galactic bulge (Mr{\'o}z \etal 2014) after resampling.
Nonetheless, the classifier was able to make similar decisions to a skilled human, indicating its efficacy. 
Future improvements could include adding new features, such as the presence of a host and WISE colors. 

The Random Forest classifier performed relatively well on known five classes.
For the other classes of transients, \eg OGLE-2012-SN-006 classified spectroscopically as Ibn (Pastorello \etal in prep.), or OGLE-2013-SN-066 which was likely an AGN flare, the classifier returned the winning class from the five trained classes, however, usually the broad probability distribution function (PDF) indicated the uncertainty of the classification.
The highest value in the PDF and corresponding class are shown in Table \tabmain~for those objects.

In the future, in order to increase the capabilities of the classifier, the training set should be expanded with more examples of cataclysmic variable outbursts and also a wider variety of Type II supernovae.
Nevertheless, we have shown that applying a very simple feature-based classification, we were able to reasonably well reproduce the classification from spectra or full light curve inspection.
This classification schema will be implemented within the OGLE-IV pipeline in future observing seasons.

\section{Supernovae Environments}
\label{sec:galfit}

The majority of the SNe from the OGLE sample are located within or close to galaxies. 
The host-galaxy properties and the supernova environment is a plausible source of inhomogeneity in SN properties (\eg Maguire \etal 2011; Childress \etal 2013), with a probable dependence on redshift since the galaxies evolve with redshift.


In order to compute the projected distances of our supernovae from the centers of their host galaxies, we first modeled the hosts' light profiles with an elliptical S\'ersic profile with S\'ersic index $n$ free to vary using the $galfit$ software (Peng \etal 2002). 
Models were obtained for hosts of 148 transients (the rest had either not present or too faint hosts) and we measured the position of the nucleus as well as an effective radius encompassing half of the total galaxy flux, denoted $R_\mathrm{Ser}$.
In Fig. \figgalfitEx~we present examples of fitted galaxies (OGLE-IV reference image, the best fit model, and the residuals).
The positions of supernovae were derived using the subtracted images from the DIA pipeline (which, by definition of the difference imaging, were registered on the same grid as the reference images), with the residuals fitted with the Gaussian profile. 
The angular offset between the supernova and its host's nucleus is listed in Table \tabmain, as well as the projected separation in kpc for cases with known distance ({\it via} redshift).
Fig. \figgalfitRall~shows the distribution of angular distances for 148 transients with successful $galfit$ models in arc seconds. 
We also added a distribution of 110 transients classified as very likely supernovae in the visual and spectroscopic classification.
The distance distribution clearly shows that OGLE-IV Transient Detection System program is finding most of its transients near or on top of the cores of the galaxies.
This is achieved thanks to superb image quality of the OGLE survey and a dedicated difference imaging software fine-tuned to the survey's data.

\begin{figure}
\centering
\includegraphics[width=1\columnwidth]{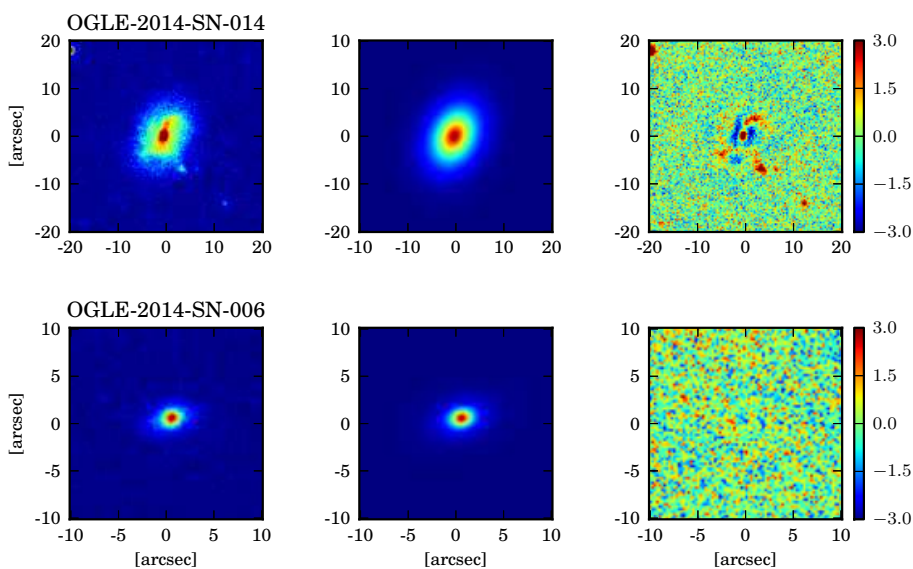}
\caption{Examples of $galfit$ models of the light of the host galaxies. The leftmost panel shows the original images from the OGLE-IV reference images, the middle are the models and the residuals are on the right. The scale of the residuals is normalized to one sigma. The upper example shows a galaxy with spiral arms, not well modeled by $galfit$, whereas the lower example is a galaxy well approximated by a single component ellipsoidal model.}
\label{fig:galfitEx}
\end{figure}

\begin{figure}
\centering
\includegraphics[width=1\columnwidth]{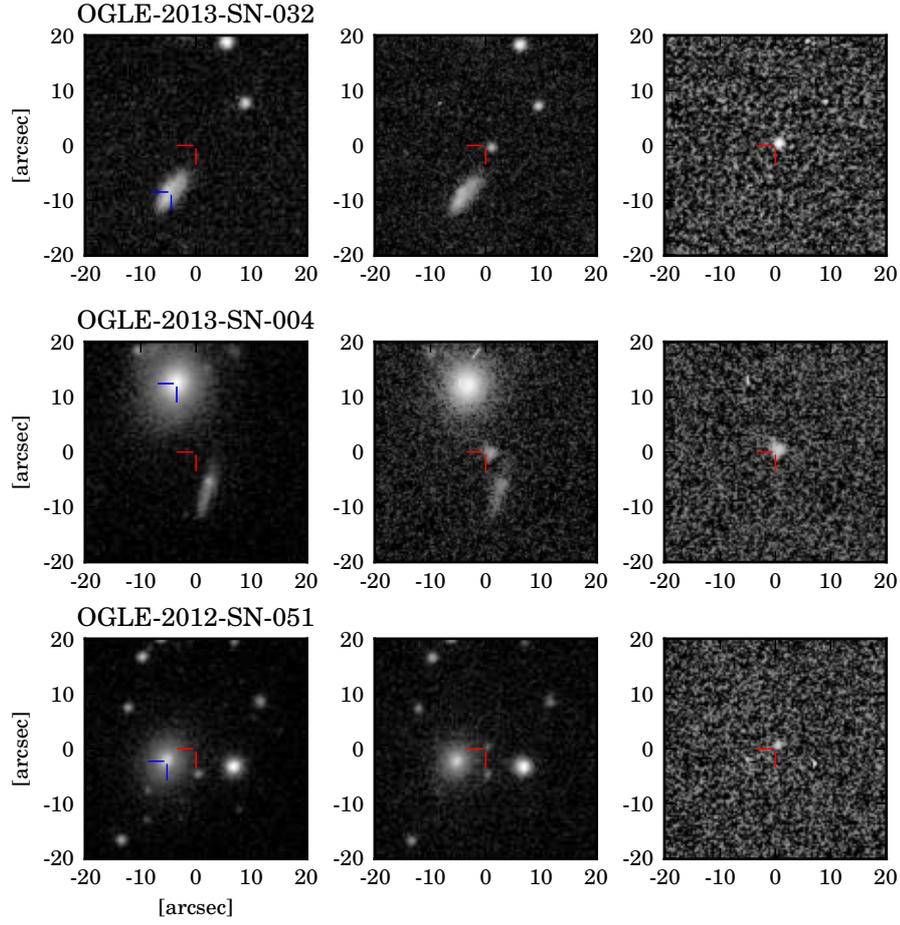}
\caption{Examples of spectroscopically confirmed supernovae separated by more than 3$R^*_\mathrm{SER}$ from the host center. From left to right: reference image, maximum brightness image, subtraction image. The red cross marks the centroid position on the subtracted image and the blue cross is the center of the galaxy from the $galfit$ model.}
\label{fig:galfitRsergtthree}
\end{figure}

Galaxies vary in morphology and size, hence in order to obtain a more homogenous picture of the distribution of supernova separations 
we normalized the distance between a SN and its host using the S\'ersic radius, taking into account the ellipticity and orientation of the galaxy.
Table \tabmain~contains the Offset$^*$, which is computed across isophotes of the galaxy light profile, such that, \eg for an edge-on galaxy, a supernova located below or above the most of the light of the galaxy will have its Offset$^*$ larger than another supernova at the same angular distance, but located along the galaxy disk.
Fig. \figgalfitRIaCC~shows examples of spectroscopically confirmed supernovae which were detected at a galactocentric distance larger than 3 S\'ersic radii computed across isophotes ($R_\mathrm{SER}^*$).
Fig. \figgalfitNohost~shows supernovae classified as Type Ia for which the host was not found within 30 arc seconds from the transient.
In most cases the host was probably too faint and was not detected on OGLE reference images (depth $\sim$22 mag, which corresponds to $M_I$<-16 mag), however in some cases the host could have been located at a much larger separation (\eg OGLE-2013-SN-006).

\begin{figure}
\centering
\includegraphics[width=1\columnwidth]{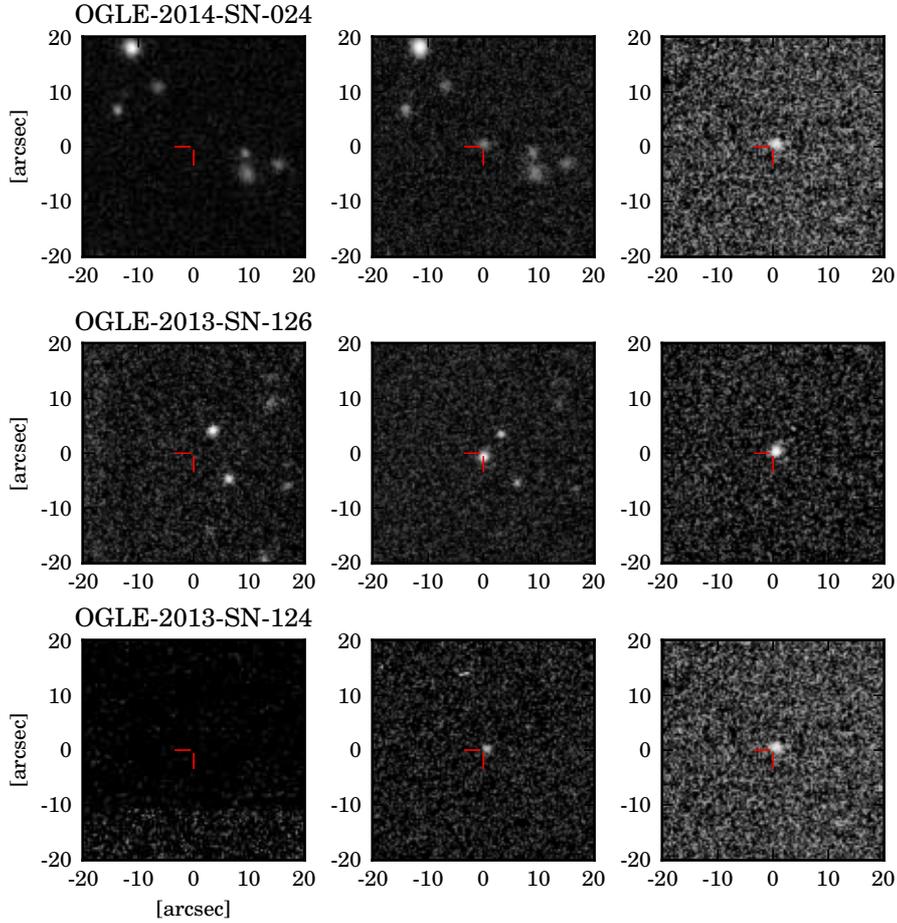}
\caption{Examples of apparently host-less supernovae with spectral confirmation as Type Ia. From the left to right: reference image, maximum brightness image, subtraction image. Red crosses mark the position of the centroid of the subtracted object.}
\label{fig:galfitNohost}
\end{figure}

\begin{figure}
\centering
\includegraphics[width=1\columnwidth]{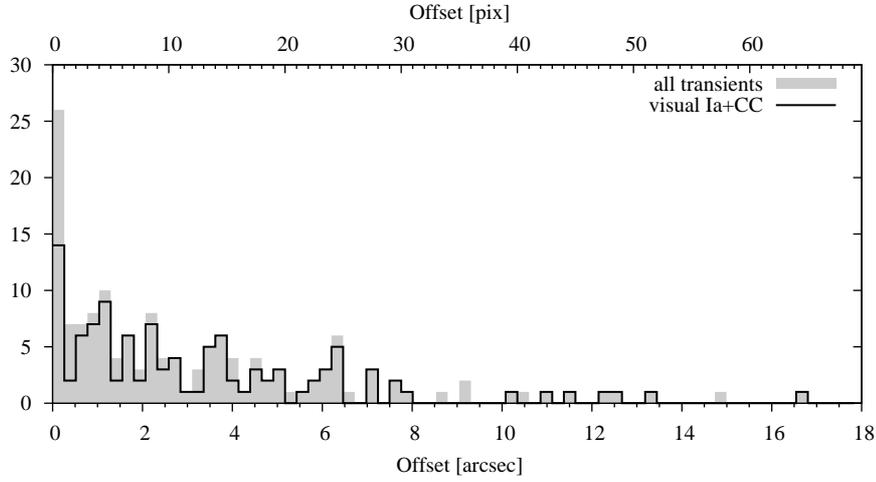}
\caption{Distribution of distance between the transient and the galaxy centre of light in arcsec and OGLE-IV pixels (0.26 arcsec) - gray: all 148 models and black line: 110 visual confirmed SNe.
About half of the transients from 26 found within radius of 1 pixel (0.26 arcsec) from galaxy center are most likely AGNs (based on spectra, previous variability or WISE colors), but we still find a significant number of SNe in the galaxy centers.
}
\label{fig:galfitRall}
\end{figure}

\begin{figure}
\centering
\includegraphics[width=1\columnwidth]{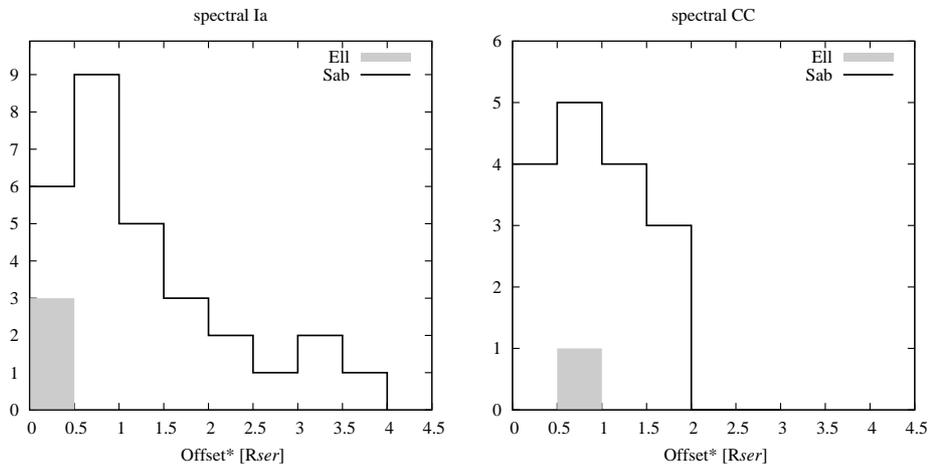}
\caption{Distribution of distance between the transient and the galaxy centre of light normalized with S\'ersic radius for: visually and spectroscopically confirmed Type Ia and CC SNe. The transients are also divided by the host galaxy type - spiral or elliptical - based on WISE colors classification.} 
\label{fig:galfitRIaCC}
\end{figure}

In the Fig. \figgalfitRall~ we present the distribution of separations between the transient and the galaxy centre in arc seconds for two samples: all transient candidates with visible hosts and transients visually and spectroscopically classified as supernovae.
After removing AGN candidates, we can see that there is still a number of transients being found in the very centers of the galaxies, what we discuss below. 

In Fig.  \figgalfitRIaCC~we show the galactocentric separations of supernovae in units of the S\'ersic radius only for spectroscopically confirmed supernovae.
We distinguish between Type Ia and CC supernovae and separate the hosts between spiral and elliptical galaxies, based on WISE colors, following the method of Assef \etal (2010) and Koz{\l}owski \etal (2013).
We notice that the majority of OGLE SNe are found in spiral galaxies.
The distribution follows the light in galaxy and half of the events lies within a single S\'ersic radius.
Deficit of supernovae below 0.5 half-light radii is mostly caused by a selection bias of the spectroscopic follow-up observations, which tend to avoid nuclear transients.

\subsection{Galaxy Orientation Bias}

Inclination plays an important role in SN detection. Spiral galaxies with higher inclination have higher surface brightness and higher extinction along the line of sight. 
Therefore, we should expect an observational bias toward finding more supernovae toward less inclined galaxies. 
We have used the galaxy short-to-long axis ratio $q=b/a$ obtained from \textit{gal fit} to compute the galaxy inclination for the sample of OGLE spiral galaxies. 
The usual Hubble (1926) formula has been used, adopting the median short-to-long axis ratio $\gamma = 0.22$ derived by Unterborn and Ryden (2008) and the correction of +3 $^{\circ}$ from Aaronson et. al.  (1980).

\begin{equation}
i = \text{cos} ^{-1} \left[ \sqrt{ \left( \frac{q^2 - \gamma^2}{ 1 - \gamma^2} \right) } \right] + 3 ^{\circ} 
\end{equation}

The resulting normalized distribution of OGLE SN hosts inclinations is shown in Fig. \figgalinclination, along with the normalized distribution of a random selection from the overall galaxy population. We observe a gap in the low inclination regime (\ie face-on galaxies).
This effect has already been observed in other surveys (Leaman et. al. 2011) and it is associated with lack of precision when measuring the major and minor axis. 
In the case of randomly oriented galaxy sample, we should expect a uniform distribution in sin($i$); however, the precision issue makes this assumption no longer valid.

To analyze the distribution of inclination angles for the SN host population, removing any possible systematic bias associated with the uncertainty in the inclination angle, we compared two populations: the first containing SN host galaxies and the second containing a random sample of galaxies (\ie contained in the OGLE field LMC571, as published in Soszy{\'n}ski \etal 2012), which have also been modeled with $galfit$. 
In order to account for spiral galaxies only, we selected only the galaxies with S\'ersic index $n < 2$ from both groups. 
The cumulative distribution of $\text{sin} (i)$ for both populations is plotted in the right panel of Fig. \figgalinclination. 
The figure, contrary to our expectations, shows an excess of objects with higher inclinations (approximately higher than 45$^\circ$) among the SN hosts, meaning that the edge-on orientation are more frequent among SN hosts. In order to quantify this effect, we run a Komolgorov-Smirnov two population test, which provides a $p$-value=0.1078. 
This result means that with a significance of 10\% we can not rule out the null hypothesis that the two populations are identical. 
The conclusion is that there is no a statistically significant host galaxy bias in our sample.

\begin{figure}
\centering
\includegraphics[width=\columnwidth]{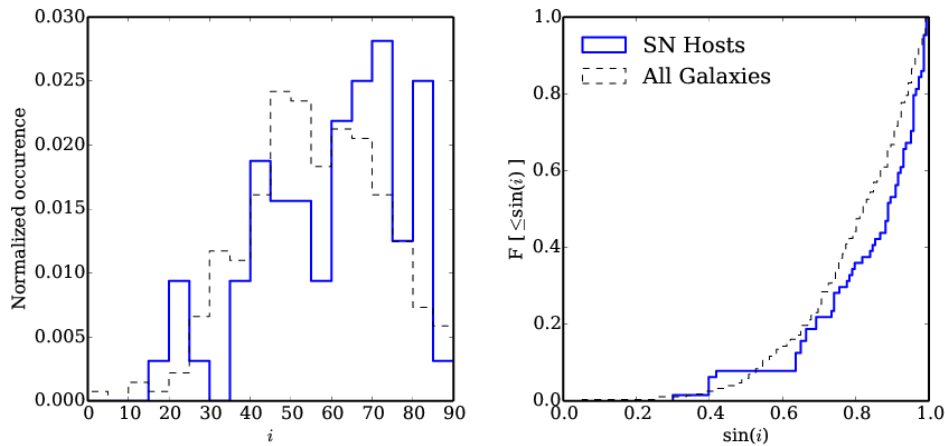} 
\caption{Left: Histogram of computed inclinations for a SN host galaxy population (blue thick line), and histogram for a random sample of spiral galaxies (thin black line). Right: Cumulative normalized distribution of sin(i) for both populations. }
\label{fig:gal_inclination}
\end{figure}

\subsection{Positional Accuracy}

The main source of uncertainty in the galacto-centric distance measurement comes from the galaxy light modeling. Positions of transients are derived from the DIA subtracted images and are typically known to better than a fraction of a pixel.
In order to assess uncertainty of the nucleus position we used a sample of AGNs, which are located at the centers of galaxies.
We selected the AGNs based on their mid-IR colors as measured by WISE (Assef \etal 2010) and picked those which also exhibited a clear variation in their light curves. 
For the ten brightest data points we extracted the mean position of the DIA residuals. 
Then, it was compared to the position of the center of the host galaxy derived using $galfit$.
We only considered nearby AGNs ($z<0.15$) where the host galaxy was clearly visible on the OGLE reference images.
For the selected sample of 16 AGNs (see Fig. \figagnsvst) the measured offsets in the $x$ and $y$ coordinates were smaller than 0.5 pixel (0.13 arc sec).
As this verification method also included the uncertainty in the transient position as measured in difference images (DIA), we find that the overall error budget for our distance determination using $galfit$ and the DIA is about 0.13 arc seconds for a typical transient of 18-19 mag.

\begin{figure}
\centering
\includegraphics{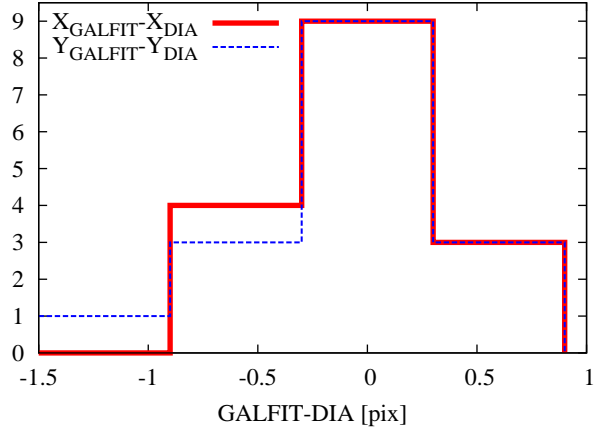}
\caption{Difference in $x$ and $y$ coordinate between $galfit$-measured center of a galaxy hosting a central AGN and a DIA-based position of the residuals at the maximum brightness of the AGN. The sample comprises of 16 AGNs, for which the $\langle\Delta x\rangle=-0.071\pm 0.355$pixels, $\langle\Delta y\rangle=-0.073 \pm 0.379$ pixels. 1 pixel is 0.26 arc seconds.}
\label{fig:agnsvst}
\end{figure}

\subsection{Transients in the Centers of Galaxies}
Interestingly, there seem to be at least a dozen of transients found within the centers of their hosts (within 1 pixel of the center) which were not classified as supernovae, neither spectroscopically nor visually.
Partially, the reason for the deficit of spectroscopically observed transients near the centers of galaxy is the natural bias of the follow-up groups, which tend to avoid taking spectra near the centers of galaxies due to centering and contamination with host galaxy light.
Nevertheless, the excess of central transients in Fig. \figgalfitRall~ is clearly visible. 
Those are probably for the most part flares or other photometric activity of AGNs, with the most obvious examples being OGLE13-088 and OGLE13-090, for which their AGN nature was also confirmed with spectroscopy.
We also used WISE color cuts to classify transients as potential AGNs following the method of Assef \etal (2010), shown in Table \tabmain~under WISE column.
Some of the central transients, however, do not seem to correspond to an AGN identified within the nucleus of the host, \eg OGLE13-066 or OGLE13-033.
Among those transients located within one pixel from the nucleus is one (OGLE13-071) with a light curve resembling that of a Tidal Disruption Event (TDE, \eg Gezari \etal 2012), but somewhat shorter with duration of only about 70 days.
More interestingly, there are also two other short events which appeared in the centers of galaxies: OGLE13-003 and OGLE13-056, which lasted for less than 40 days, see Fig. \figLCnuclear~ for their light curves.
It is currently difficult to conclude upon their nature with no spectroscopic follow-up, however, we can expect similar transients being found in the OGLE data in the following seasons, and hope for real-time detections of TDEs and other exotic nuclear transients, allowing their detailed studies.

\begin{figure}[tb]
\centering
\includegraphics[width=1\columnwidth]{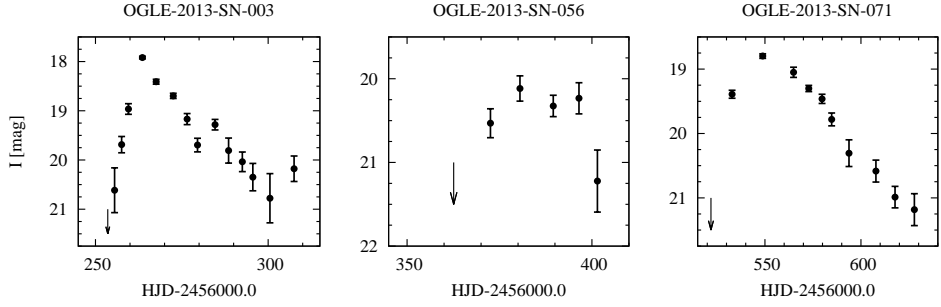}
\caption{Nuclear transients of not obvious nature found within 1 pixel (0.26 arcsec) from the cores of their hosts.}
\label{fig:LC1}
\end{figure}

\begin{figure}[b]
\centering
\includegraphics[width=1\columnwidth]{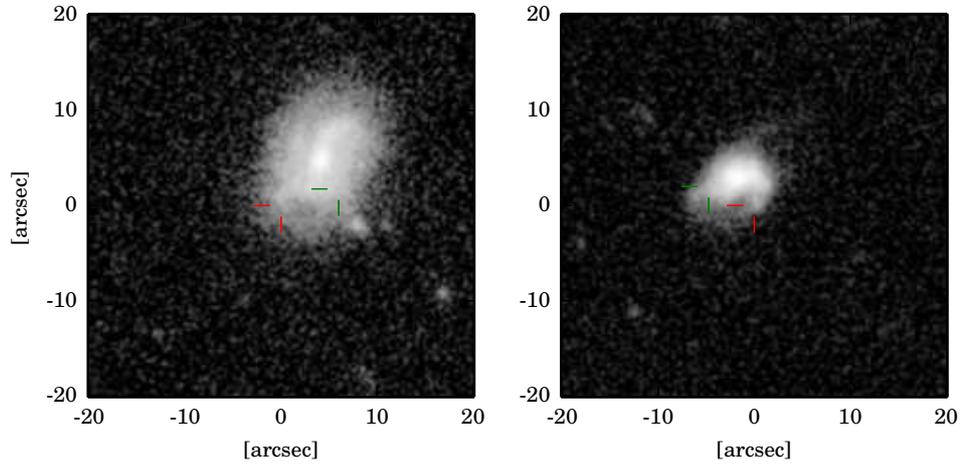}
\caption{Supernova factories: galaxies with two supernovae found within one and two years, respectively. Left: Type IIn OGLE-2013-SN-017 (green) and type Ib OGLE-2014-SN-014 (red). Right: OGLE-2011-SN-034 (green) (from Koz{\l}owski \etal 2013) and OGLE-2013-SN-022 (red), both of unknown types.}
\label{fig:pairs}
\end{figure}

\subsection{Supernova Factories}
We note that among all OGLE transients, there were two pairs of supernovae which exploded in the same host galaxy.
Such cases are important for studying supernovae environments and the supernova rates (\eg Th{\"o}ne \etal 2009).
The first pair, OGLE-2013-SN-017 and OGLE-2014-SN-014, appeared clearly in the same host, the galaxy 2MASX J04272268-7442059 (after NED), separated by 6.36 arcseconds, \ie 5.3 kpc at z=0.043. 
SN OGLE13-017 was classified as a Type IIn (Inserra \etal 2013), and the other one, which exploded about a year later, SN OGLE14-014 was also core-collapse, type Ib (Marion \etal 2014).
The second pair, OGLE-2011-SN-034 and OGLE-2013-SN-022, consists of an archival (Koz{\l}owski \etal 2013) and real-time detections.
The separation in this case was 5.22 arcsec, however, neither of those supernovae were classified, nor we know the redshift of the host galaxy (GALEXASC J025017.60-705225.4 after NED). 
Their light curves are not completely covered, therefore their photometric classification is also difficult, however, most likely they both are SNe Type Ia. 
The reference images with hosts and positions of paired supernovae are shown in Fig. \figpairs.
We note, that there is also an apparent pair: OGLE-2013-SN-034 and OGLE-2013-SN-055, which are the same supernova, but given two separate IDs by mistake.

\section{Cosmology with OGLE Supernovae}
A key goal of most supernova surveys is detecting Type Ia supernovae, which are known to be ``standardizable candles'', and hence can be used for cosmological studies of the expansion of the Universe (\eg Riess \etal 1998, Perlmutter \etal 1999, Sullivan \etal 2011, Campbell \etal 2013).
Our sample from the OGLE-IV survey consists of 49 spectroscopically confirmed Type Ia SNe within the redshift $z<0.14$ with a median value of z=0.076.

The absolute magnitude of the SN observed at magnitude $m$ is described by the equation:
\begin{equation}
M=m-\mu-A_{\mathrm{MW}}-A_{\mathrm{H}}-K(z),
\end{equation}
where $\mu$ is the distance modulus, $A_{\mathrm{MW}}$ and $A_{\mathrm{H}}$ are the extinctions in the Milky Way and in the SN host galaxy in the $I$-band, respectively.
The single filter $K$-correction, which accounts for brightness difference due to redshifted spectrum, was adopted from the Magellanic Bridge supernova sample in the OGLE-IV (Koz{\l}owski \etal 2013).
The distance moduli to the SNe were calculated from spectroscopic redshifts assuming the $\Lambda$CDM cosmological model from the Planck mission with parameters: $H_0=68$ km/s/Mpc, $\Omega_M=0.31$, $\Omega_\Lambda=0.69$ (Planck collaboration 2013).
We took into account the Milky Way extinction toward each SN using Galactic Extinction maps from Schlafly and Finkebeiner (2011)
(retrieved {\it via} the NASA/IPAC Extragalactic Database).
Due to lack of color light curves we were not able to fit the color for our light curves, hence the host galaxy extinction remained as the only unknown parameter.

The distance moduli were also derived from the light curve fitting and here we used the empirical method for fitting multi-color light curves  of Type Ia SNe described by Prieto \etal (2006).
This method relies on the calibrated relation between the absolute magnitudes at maximum light and the post maximum decline rate $\Delta \mathrm{m_{15}}$ (brightness change from maximum to 15 days post maximum) in $BVRI$ filters.
By fitting the parameter $\Delta \mathrm{m_{15}}$ and using the linear relation between the absolute magnitude at maximum and the post maximum decline rate we obtained the absolute magnitude, and hence the distance moduli for our supernovae.
For fitting we used the well sampled $I$-band light curves and, where available, $V$-band data.
In Fig.  \figIaPrieto~we show examples of OGLE Type Ia supernovae and the best fitted template from Prieto \etal (2006).

\begin{figure}
\centering
\includegraphics[width=1\columnwidth]{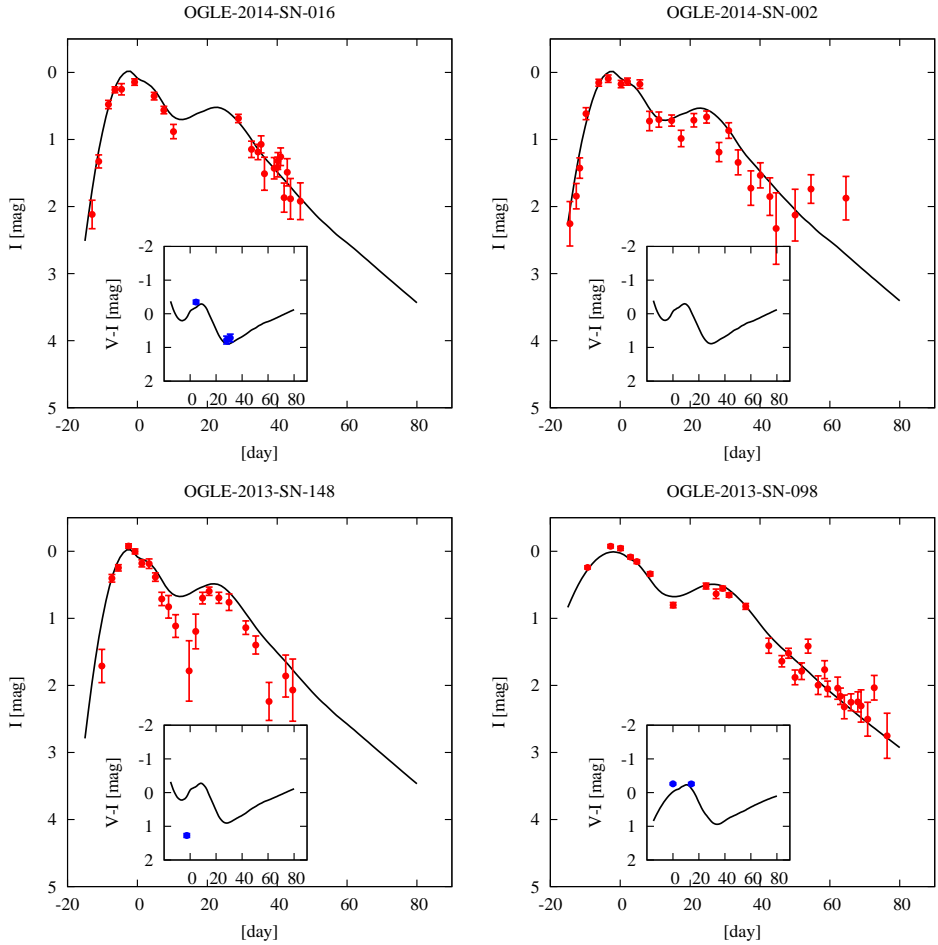}
\caption{Example light curves of Type Ia supernovae from OGLE-IV along with their models from Prieto \etal (2006) (black solid line). The time axis shows days from the $I$-band maximum. The inset in each panel shows the available $V-I$ data with the model. 
Supernova OGLE-2013-SN-148 exhibits a very deep dip between the two peaks and its model is poorly fitted. Also its color around maximum is significantly redder than other supernovae, indicating a significant amount of extinction.}
\label{fig:IaPrieto}
\end{figure}

In Fig. \figIaHubble~we present the Hubble Diagram for 49 Type Ia SNe from the OGLE sample with spectroscopic redshifts.
The final error in the distance modulus is obtained by adding in quadrature the instrumental error in measured magnitude, the scatter of fitting SNe templates, and the dispersion in $\Delta m_{15}$ (Prieto \etal 2006).
We compared our data to the $\Lambda$CDM cosmology model with Planck parameters (Planck collaboration, 2013) and obtained the scatter of 0.315 mag in residuals.
Table \tabIaparam~presents the results of the $\Delta m_{15}$ fits to the light curves of 49 Type Ia SNe in the OGLE sample with the columns with $\Delta m_{15}$ value, the distance modulus derived from $\Delta m_{15}$, residuals on the Hubble diagram for the Planck cosmological models, the isophotal offset between a SN and the galaxy center (where the light model was available) in units of the S\'ersic radius.

The Hubble Diagram residuals exhibit relatively low scatter of 0.315 mag, despite the fact that the results relied on single-band light curves. 
However, our light curves were in most cases well covered from before the maximum until the supernova disappeared, allowing for good template fitting.
The residuals exhibit clearly a systematic positive offset, which is expected, as we did not include any host galaxy extinction in the distance moduli calculations.
Ignoring other more subtle effects on the residuals, like the host mass or metallicity, we can therefore use them to infer the amount of the host extinction for each of the supernovae. 
The most striking outlier is supernova OGLE13-148, which exhibits the most significant deviation from the expected brightness by more than 1 mag.
Also, a single $V-I$ measurement near the first peak indicates the host extinction was very large in this case, but the finding chart shows that this supernova appeared on top of a well-pronounced spiral arm of a large galaxy.
On the other hand, as seen in Fig.  \figIaPrieto, the model of the light curve of OGLE13-148 matches well everywhere except for the dip between the two $I$-band peaks.
However, the significant dip by more than 1 mag is hard to explain even by somewhat higher extinction, and this object requires more detailed studies.

In Fig.  \figIaOffset~we show the offsets between supernovae locations and their host galaxy centers against their residuals on the Hubble Diagram.
We notice a subtle increase in the residuals while approaching the normalized center of the host galaxy, however, after about two half-light radii the residuals tend to agree with zero.
Ignoring the larger scatter in the residuals below 0.1 $R_{Ser}$ and a few negative residuals, likely due to inaccuracies in galaxy core subtraction, we can see that in range from 0.1 to 2 $R_{Ser}$ the systematic offset can be approximated as constant. 
Assuming it is all caused by the host extinction we can derive a mean value of the extinction for this range of normalized galactocentric separations of  $A_I=0.19\pm0.10$ mag.
Based on the relation between $A_I$ and $A_V$ from Rieke and Lebofsky (1985) we derive $A_V=0.39 \pm 0.21$ mag, in agreement with 
Galbany \etal (2012) who obtained the mean value of $A_V=0.36\pm 0.02$ mag for the SDSS sample of Type Ia SNe in spiral galaxies.
They have also derived a linear dependence of extinction on the projected distance from the host galaxy center, however, we do not see this trend in our data. 

For OGLE-IV Type Ia supernovae located between 0.1 and 2 half-light radii from the host centers we used the derived correction for the extinction and for the rest we assumed no or negligible extinction.
Hubble Diagram with the Union 2.1 sample of supernovae (Suzuki \etal 2012) and 49 supernovae from OGLE-IV is shown in Fig. \figHubbleUnion. 
Except for a couple of clear outliers due to very high extinction or bad models, our ensemble of supernovae fits well within the ``gold-sample'' of cosmologically useful supernovae Type Ia, complementing the sample in the least populated redshift range from 0.06 to 0.11.

\begin{figure}
\centering
\includegraphics[width=1\columnwidth]{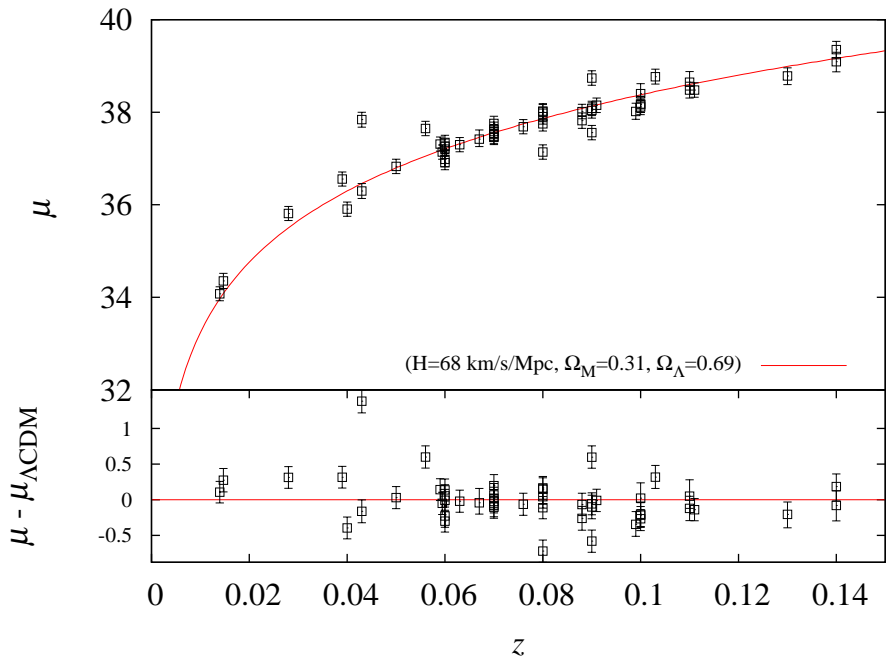}
\caption{
Top: Hubble diagram for the OGLE sample:
the distance modulus--redshift relation ($\mu_{\Lambda \mathrm{CDM}}$) of the assumed $\Lambda$CDM cosmological model from Planck.
Bottom: Residuals from the assumed $\Lambda$CDM cosmological model as a function of redshift. 
Mean offset and its $rms$ is $0.004\pm0.315$.}
\label{fig:IaHubble}
\end{figure}

\begin{figure}
\centering
\includegraphics[width=1\columnwidth]{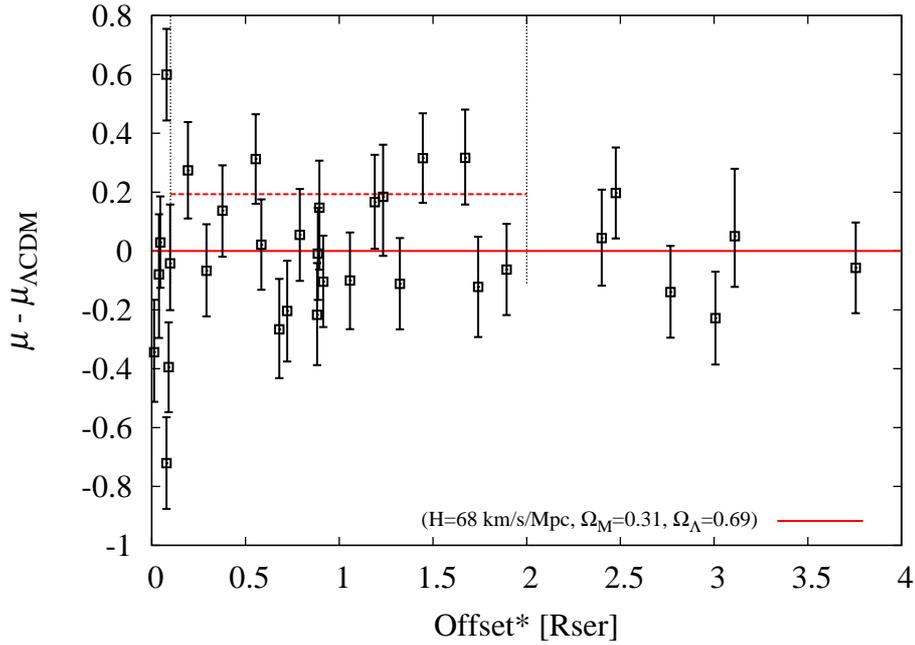}
\caption{Residuals on the Hubble Diagram as a function of the offset between the SN location and the host galaxy nucleus in units of S\'ersic radius.
We show the results for 33 Type Ia SNe for which the hosts were detected and successfully modeled with $galfit$.
The host extinction in the range between 0.1 and 2 half-light radii computed from the mean residuals is $A_{I}=0.19 \pm 0.10$ mag (dashed line) assuming Planck cosmological parameters. Central transients (distance below 0.1) exhibit significant scatter in residuals most likely due to inaccuracies in galaxy core subtraction.}
\label{fig:IaOffset}
\end{figure}

\begin{figure}
\centering
\includegraphics[width=1\columnwidth]{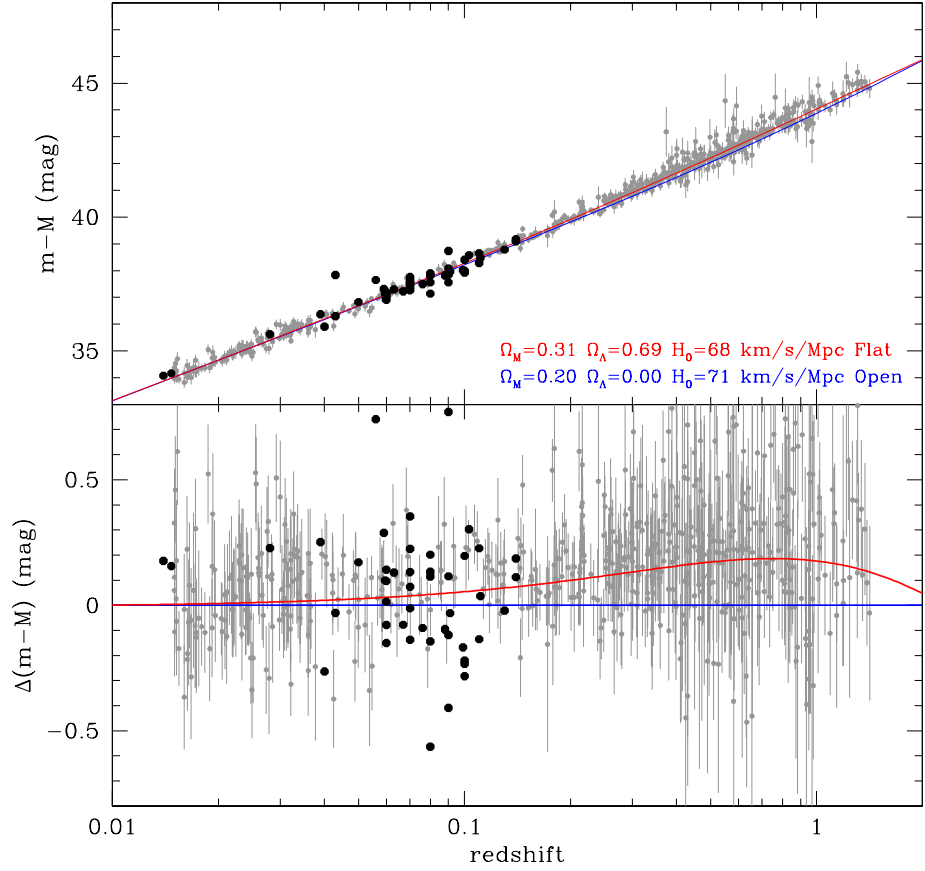}
\caption{
Hubble Diagram with OGLE-IV (black dots) and Union 2.1 sample (gray) of supernovae Type Ia along with Planck (flat) and open cosmological model.
OGLE-IV supernovae were corrected for extinction based on their distance from the host center.
}
\label{fig:HubbleUnion}
\end{figure}

\begin{table}
\footnotesize
\begin{tabular}{c l c c c c c}
\hline
ID & z & A$_\mathrm{MW}$ & $\Delta \mathrm{m_{15}}$ & $\mu$ & $\mu-\mu_{\Lambda \mathrm{CDM}}$ & Offset* \\
OGLE-20.. & & [mag] & & [mag] & [mag] &  [R$_\mathrm{Ser}$]\\
\hline
14-SN-024  &  0.1  &  0.244  &  0.839  &  38.403  & 0.028 &  \\ 
14-SN-021  &  0.039  &  0.113  &  1.183  &  36.556  & 0.321 &  1.45 \\
14-SN-019  &  0.04  &  0.113  &  1.260  &  35.905  & -0.39 &  0.09 \\
14-SN-016  &  0.07  &  0.287  &  1.299  &  37.536  & -0.019 &  \\ 
14-SN-010  &  0.056  &  0.109  &  1.247  &  37.649  & 0.604 &  0.08 \\
14-SN-002  &  0.10  &  0.063  &  1.289  &  38.114  & -0.261 &  0.68 \\
13-SN-156  &  0.14  &  0.055  &  1.234  &  39.090  & -0.074 &  0.04 \\
13-SN-148  &  0.043  &  0.150  &  1.368  &  37.841  & 1.386 &  \\ 
13-SN-147  &  0.099  &  0.113  &  1.281  &  38.016  & -0.339 &  0.01 \\
13-SN-141  &  0.05  &  0.189  &  1.307  &  36.829  & 0.034 &  0.05 \\
13-SN-136  &  0.080  &  0.065  &  0.860  &  38.026  & 0.171 &  1.19 \\
13-SN-130  &  0.09  &  0.095  &  1.052  &  38.738  & 0.603 &  \\ 
13-SN-129  &  0.08  &  0.028  &  1.337  &  37.904  & 0.049 &  2.40 \\
13-SN-126  &  0.06  &  0.051  &  1.246  &  37.202  & -0.003 &  \\ 
13-SN-124  &  0.13  &  0.065  &  1.191  &  38.785  & -0.2 &  \\ 
13-SN-123  &  0.08  &  0.078  &  1.457  &  37.139  & -0.716 &  0.08 \\
13-SN-120  &  0.07  &  0.113  &  1.644  &  37.628  & 0.073 &  \\ 
13-SN-118  &  0.07  &  0.113  &  0.882  &  37.757  & 0.202 &  2.48 \\
13-SN-109  &  0.088  &  0.036  &  1.096  &  38.013  & -0.062 &  0.29 \\
13-SN-099  &  0.028  &  0.478  &  0.953  &  35.812  & 0.317 &  0.55 \\
13-SN-098  &  0.06  &  0.044  &  1.068  &  37.347  & 0.142 &  0.38 \\
13-SN-096  &  0.11  &  0.087  &  1.158  &  38.478  & -0.117 &  1.74 \\
13-SN-080  &  0.103  &  0.076  &  1.096  &  38.766  & 0.321 &  1.67 \\
13-SN-075  &  0.08  &  0.038  &  1.117  &  37.748  & -0.107 &  1.32 \\
13-SN-073  &  0.091  &  0.057  &  1.257  &  38.151  & -0.004 &  0.89 \\
13-SN-070  &  0.043  &  0.029  &  1.734  &  36.297  & -0.158 &  \\ 
13-SN-057  &  0.10  &  0.032  &  1.265  &  38.163  & -0.212 &  0.88 \\
13-SN-051  &  0.07  &  0.024  &  1.107  &  37.456  & -0.099 &  0.92 \\
13-SN-044  &  0.07  &  0.077  &  1.069  &  37.581  & 0.026 &  0.58 \\
13-SN-043  &  0.14  &  0.031  &  1.025  &  39.354  & 0.189 &  1.23 \\
13-SN-041  &  0.10  &  0.034  &  1.192  &  38.176  & -0.199 &  0.72 \\
13-SN-040  &  0.09  &  0.033  &  1.219  &  38.039  & -0.095 &  1.06 \\
13-SN-039  &  0.08  &  0.056  &  1.132  &  38.007  & 0.152 &  0.89 \\
13-SN-032  &  0.09  &  0.035  &  1.097  &  38.083  & -0.052 &  3.76 \\
13-SN-018  &  0.067  &  0.113  &  1.441  &  37.418  & -0.037 &  0.10 \\
13-SN-015  &  0.088  &  0.051  &  1.714  &  37.819  & -0.257 &  \\ 
13-SN-009  &  0.060  &  0.060  &  1.243  &  37.265  & 0.06 &  0.79 \\
13-SN-004  &  0.06  &  0.140  &  1.415  &  36.982  & -0.223 &  3.01 \\
13-SN-001  &  0.09  &  0.026  &  1.266  &  37.559  & -0.576 &  3.11 \\
12-SN-051  &  0.11  &  0.045  &  1.482  &  38.650  & 0.055 &  \\ 
12-SN-049  &  0.07  &  0.056  &  1.303  &  37.477  & -0.078 &  2.77 \\
12-SN-046  &  0.111  &  0.050  &  1.182  &  38.480  & -0.135 &  \\ 
12-SN-044  &  0.06  &  0.034  &  1.182  &  36.910  & -0.295 &  \\ 
12-SN-040  &  0.014690  &  0.113  &  1.889  &  34.354  & 0.279 &  0.19 \\
12-SN-032  &  0.063  &  0.113  &  1.216  &  37.299  & -0.016 &  \\ 
12-SN-014  &  0.059  &  0.113  &  1.069  &  37.312  & 0.147 &  \\ 
12-SN-009  &  0.013966  &  0.046  &  1.063  &  34.076  & 0.111 &  \\ 
12-SN-007  &  0.059438  &  0.103  &  1.293  &  37.139  & -0.046 &  \\ 
12-SN-005  &  0.076  &  0.042  &  1.207  &  37.687  & -0.058 &  1.89 \\
\\
\hline
\end{tabular}
\caption{Parameters of the 49 Type Ia supernovae modeled using Prieto \etal (2006) templates. 
$A_\mathrm{MW}$ is the Milky Way extinction in the $I$-band. $\Delta m_{15}$ is the parameter obtained from the template fits, $\mu$ is the derived distance modulus, $\mu-\mu_{\Lambda CDM}$ are the HD residuals for different cosmologies and Offset$^*$ is the galacto-centric distance of a supernova in units of the Sersic radius (only for hosts with good $galfit$ models).}
\label{tab:Iaparam}
\end{table}

\section{Summary}
Since 2010, the OGLE-IV survey has been annually discovering $\sim$ 150 transients in the regions of the sky around the Magellanic Clouds. 
The OGLE-IV Transient Detection System surveys 650 deg$^2$ of the sky to a depth of $I<21$mag from 2012 and provides an unbiased sample of all types of supernovae to $z\sim$0.15.
Excellent data quality provided by the state-of-the-art fine-tuned difference imaging delivers high quality and well sampled light curves, as well as discoveries from a range of galactic environments, including both very dense central and remote locations.

We presented the data for all 238 transients found in real-time in years 2012-2014, among which a significant fraction were classified spectroscopically. 
Supernovae Type Ia were used to construct the Hubble Diagram with relatively small intrinsic scatter. 
Systematics in the residuals were also used to derive the mean value of host extinction in the range up to two half-light radii. 

Projected galactocentric distances of most of the transients were measured, and we showed that OGLE-IV is capable of detecting nuclear transients with good efficiency. We presented a few examples of interesting central transients of unknown nature.

The OGLE-IV Transient Detection System is expected to continue in the future, extending the sample of well covered supernovae light curves and providing more interesting cases for detailed studies.

\Acknow{
We thank Drs M. Fraser, D. Poznanski, H. Campbell for their continued support and help in preparation of this manuscript. 
We would like to express our gratitude to the members of the spectroscopic follow-up groups, especially the PESSTO team, CSP (in particular to E. Y.  Hsiao, G. H. Marion and M. Phillips) and M. Childress (ANU).

We also would like to thank the fellows of the Polish Children's Fund, Jakub Aniulis, Jakub Banaszak and Jakub Mnich, for their involvement in this work.

This work made use of observations collected at the European Organisation for Astronomical Research in the Southern Hemisphere, Chile as part of PESSTO, (the Public ESO Spectroscopic Survey for Transient Objects Survey) ESO program ID 188.D-3003.

This research has made use of the NASA/IPAC Extragalactic Database (NED) which is operated by the Jet Propulsion Laboratory, California Institute of Technology, under contract with the National Aeronautics and Space Administration.

This publication makes use of data products from the Wide-field Infrared Survey Explorer (WISE), which is a joint project of the University of California, Los Angeles, and the Jet Propulsion Laboratory/California Institute of Technology, funded by the National Aeronautics and Space Administration.


The OGLE project has received funding from the European Research Council
under the European Community's Seventh Framework Programme
(FP7/2007-2013)/ERC grant agreement no. 246678 to AU. This work has been
supported by the Polish Ministry of Science and Higher Education through
the program ``Ideas Plus'' award No. IdP2002 000162 to IS.
}

\end{document}